\documentstyle[aps,prb,epsf,psfig,eqsecnum]{revtex}

\newcommand{\beq}{\begin{equation}}
\newcommand{\eeq}{\end{equation}}
\newcommand{\bea}{\begin{eqnarray}}
\newcommand{\eea}{\end{eqnarray}}
\newcommand{\nn}{\nonumber}

\newcommand{\eps}{\epsilon}
\newcommand{\veps}{\varepsilon}
\newcommand{\th}{\theta}
\newcommand{\al}{\alpha}
\newcommand{\s}{\sigma}
\newcommand{\lam}{\lambda}
\newcommand{\de}{\delta}

\newcommand{\be}{\beta}

\newcommand{\ra}{\rangle}
\newcommand{\la}{\langle}

\newcommand{\ga}{\gamma}

\newcommand{\app}{\approx}
\newcommand{\ua}{\uparrow}
\newcommand{\da}{\downarrow}
\newcommand{\Ua}{\Uparrow}
\newcommand{\Da}{\Downarrow}
\newcommand{\dmi}{{1\over 2}}
\begin{document}
\setcounter{page}{1}
\topmargin 0pt
\oddsidemargin 5mm
\title{Persistent currents through  a quantum dot} 
\author{Pascal Simon$^{1}$ and Ian Affleck$^{1,2}$} 
\address{$^1$Department of Physics and Astronomy and 
$^2$Canadian Institute for Advanced Research , University 
of British Columbia, Vancouver, BC, V6T 1Z1, Canada } 
 \maketitle 
\begin{abstract} 
We study the persistent currents induced by the Aharonov-Bohm effect
in a closed ring which either embeds  or is directly 
side coupled to a   quantum dot at Kondo resonance.
We predict that in both cases, the persistent current is very sensitive
to the ratio between the length of the ring and the size of the
Kondo screening cloud which appears as a fundamental prediction of 
scaling theories of the Kondo effect. Persistent current measurements
provide therefore  an opportunity to detect this 
cloud which has so far never been observed experimentally.
\end{abstract}

\section{Introduction}
The Kondo effect has become  one of the most studied 
paradigms in condensed matter theory in decades.\cite{Hewson}
The effect  results
from the interaction between conduction electrons and localized
magnetic impurities. 
As a paradigm, the Kondo effect has played a considerable role 
in the development of renormalization group (RG) scaling ideas.
The Kondo effect is associated with a large distance scale $\xi_K\app \hbar v_F
/T_K$, where $v_F$ is the Fermi velocity and $T_K $, the Kondo temperature, which also defines the energy scale of the problem: $T_K\sim De^{-1/\lam}$, 
where $D$ is the bandwidth and $\lam$ the bare dimensionless Kondo coupling.
 The heuristic 
picture associated with this fundamental length scale is that a cloud of electrons with a size of order $\xi_K$ surrounds the impurity, forming a singlet with it. The remaining electrons outside the cloud do not feel  the impurity spin any more but rather a  scattering potential, 
caused by the complex formed between 
the cloud and the impurity, resulting in a $\pi/2$ phase shift
at the Fermi energy. According to the renormalization group approach to the
Kondo problem, all physical quantities should be expressed in terms 
of universal scaling functions of $r/\xi_K$, r being the distance to impurity
.\cite{Sorensen,Barzykin}
Unfortunately,  this large length scale of order $0.1$ microns in metals 
has never been detected experimentally.

Quite recently, due to some experimental breakthroughs, various aspects
of the Kondo effect
have been measured in a semiconductor quantum dot coupled via weak 
tunnel junctions to leads and capacitively to gates.
By tuning the gate voltage, the number of electrons inside the quantum dot can be adjusted due to the Coulomb blockade. In particular, when the number of electrons is odd, the quantum dot can act as a localized spin $S=1/2$ impurity.
Nevertheless, unlike magnetic impurities in metals
the physical parameters of the quantum dots can be varied continuously.
When the tunneling matrix elements between the leads and the quantum dot are 
small, the system can be essentially described by a Kondo model.
It has been predicted theoretically for a Kondo impurity embedded between two 
leads, that
the transmission probability should reach one at low temperature $T<T_K$
whereas it should be small at higher temperature.\cite{Glazman,Ng} Such a manifestation of the Kondo effect has been confirmed experimentally
in the last few years.\cite{Goldhaber,Cronenwett,Simmel,Wiel}

The Kondo temperatures in these experiments are generally 
  considerably smaller than $1^0K$ and  most importantly can
be tuned via the gate 
  voltage $V_g$. Therefore,  the Kondo length scale is expected to be 
  of order 1 micron or larger. This new experimental realization of the 
Kondo effect seems  therefore to 
offer new opportunities to measure the screening cloud. 
In [\onlinecite{letter}],
we have emphasized that the transmission probability of a quantum dot, 
 in the Kondo regime,
and embedded in a closed ring, may be sensitive to the length of the ring 
versus
the Kondo screening length. The idea is simply that a finite size should suppress the Kondo effect even if the temperature is much less than $T_K$.
Such an idea, central to the interpretation of the Kondo problem, has been
for example used by Nozi\`eres to interpret Wilson's numerical renormalization group approach of the Kondo problem \cite{Nozieres}.

\begin{figure}[b]
\psfig{figure=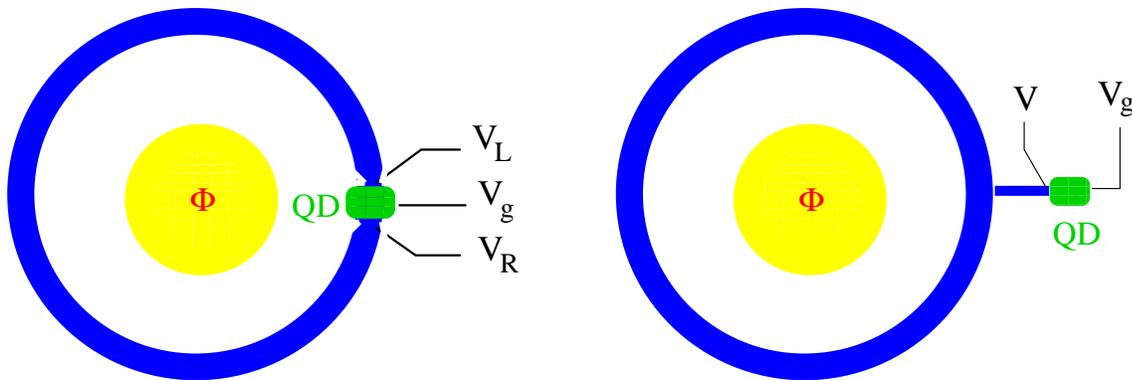,height=5cm,width=15cm}
\caption{Left: Embedded quantum dot. Right: side coupled quantum dot.
$V_g$ is the gate voltage and $V_{R/L},~ V$ control the tunneling amplitude
between the quantum dot and the ring.}
\label{dots}
\end{figure}

The purpose of this paper  is to analyze in more detail the sensitivity
of  the transmission probability of the quantum dot to the Kondo length scale 
in two different geometries: one where the
quantum dot is embedded in a closed mesoscopic circular 
ring \cite{Buttiker,Ferrari,Kang}, and another one where the quantum dot 
is outside the ring 
and couples directly to it.\cite{Buttiker,Eckle,Cho} We have shown schematically
 both devices
in figure \ref{dots}.  In both geometries, the crucial point is that 
the screening cloud is ``trapped'' in 
the ring and cannot escape into macroscopic leads. 
A natural way to measure the transmission probability through a quantum dot in a closed geometry  would be by persistent current measurements.
One might expect intuitively that the 
   persistent current, as a function of the flux, $\Phi$ penetrating the 
   ring, will be much different when the screening cloud is small 
   compared to the circumference, $l$, of the ring than when it is much 
   larger.
Such persistent current experiments have been performed 
   recently on micron sized rings not containing a quantum dot 
\cite{Chand,Mailly}. 
We  have shown that, when $\xi_K/l<<1$, the 
   persistent current is that of a perfect ring with no impurity ($j\propto ev_F/l$)
for the embedded quantum dot and is vanishing for the side coupled quantum dot. 
  On the other hand, when $\xi_K/l>>1$, $jl$ becomes much 
   smaller in the first geometry, vanishing as a power of the bare Kondo coupling, whereas it converges toward that of a perfect ring in the second geometry.  We always assume in this article that $l\gg a$, with $a$ the lattice 
constant. In 
   general, we expect $j(\Phi )l$ to be a universal scaling function of 
   $\xi_K/l$, in the usual scaling limit of the Kondo model (i.e. 
   at small Kondo coupling and large ring size compared to the lattice 
 constant).

The plan of the article is as follows. In section 2, 
we first introduce the two tight binding models we are to study
and develop a continuum limit analysis of the models in an open geometry
which clarifies the difference between the two models.
In section 3, we calculate  
the persistent currents for the embedded quantum
dot in the Kondo regime in both limits $\xi_K\gg l$ and $\xi_K\ll l$,
and discuss the extension of these results for asymmetric tunneling
amplitude between the quantum dots and the wires. A special emphasis on the role played by particle hole symmetry is also placed. We also study in this section how bulk interactions modifies the conductance of the quantum dot and hence
the persistent current.
In section 4, we present similar calculations for the side coupled quantum dot. Finally, in section 5, we discuss our results and the approximations we have used. Some technical details about our  calculations have been  relegated 
to three appendices.
  
\section{Formulation of the  models}
The first model we are considering corresponds to a quantum dot embedded
in a wire.
In order to describe this situation,       
we begin with the simple tight-binding Anderson model: 
   \begin{equation} 
   H_1=-t\sum_{j\leq -2}(c^\dagger_jc_{j+1}+h.c.)-t\sum_{j\geq 1} 
   (c^\dagger_jc_{j+1}+h.c.) 
   -t'[c^\dagger_d(c_{-1}+c_1)+h.c.] 
   +\epsilon_dc^\dagger_dc_d 
   +Un_{d\uparrow}n_{d\downarrow}.\label{Hand}\end{equation} 
   Here sums over the electron spin index are implied.  
   The quantum dot denoted by the subscript $d$  simply corresponds 
to the site $0$ of the 
   chain.  $t'$, $\epsilon_d$ and $U$ are the tunneling matrix 
   elements, gate voltage and charging energy of the dot, 
   respectively.  $n_{d\alpha}$ is the electron number at the origin 
   for  spin $\alpha$.  

The second model we want to consider corresponds to a situation where  the quantum dot is outside of the ring. It has been referred to as the side coupled
 quantum dot.
This  can be modeled by the following tight-binding Anderson model:
\beq
H_2=-t\sum\limits_{i=-\infty}^{+\infty} (c_{i+1}^\dag c_{i} +h.c.)-t'[ c^\dag_d c_0+c^\dag_0 c_d]
+\eps_d c^\dag_d c_d +Un_{d\ua}n_{d\da}~,
\label{and}
\eeq
with similar notations and conventions.
In this model the quantum dot couples directly to the  site $0$.

In order to distinguish both models, we will use in the following 
the notations $1$ for the embedded quantum dot (EQD) and $2$ 
for the side coupled quantum dot (SCQD). 
These two models have already received some attention in the past.
The EQD has been first studied in [\onlinecite{Glazman}] and
[\onlinecite{Ng}]. In particular, Ng and Lee \cite{Ng} have shown using Langreth formula \cite{Langreth} that   the conductance at zero temperature (and $t'\ne 0$) 
reads: 
\beq
G_1={2e^2\over \hbar}\sin^2\delta,
\label{glee}
\eeq
with $\delta$ the phase shift which has been  approximated by:  
\beq
\delta={\pi\over 2}\la n_d\ra,\label{delta}
\eeq
 and $\la n_d\ra$ the average occupation number of the quantum dot. Nevertheless, we would like to stress that the phase shift according to the generalization of the Friedel sum rule by Langreth \cite{Langreth} is 
proportional
to the number of  electrons displaced by the impurity ``among
which are included not only the $d$ electrons, but also some of the conduction
electrons''. The approximation of Eq. (\ref{delta}) should be valid only when the reservoir bandwidth is much larger than the bare Kondo coupling (proportional to $t'^2$) which implies that the contribution of the conduction electrons to the
phase shift is negligible. We will come back to this point later (see section
\ref{phb}) with
a special emphasis on the  role played by particle-hole symmetry.
Kang {\it et al.} \cite{Kang} have proved using similar reasoning
that the conductance of the SCQD  can be expressed as:
\beq
G_2={2e^2\over \hbar}\cos^2\delta,
\label{gkang}
\eeq
with similar notations.

In both models, the Kondo regime corresponds to the limit: 
   \begin{equation} 
   t'<< -\epsilon_d, U+\epsilon_d ,\end{equation} 
   ($\epsilon_d<0$). 
   In this limit the dot essentially is locked into the singly 
   occupied state and thus corresponds to an $S=1/2$ spin 
   impurity. In this regime, the EQD is predicted to exhibit perfect conductance according to Eq. (\ref{glee}) whereas the SCQD should give  zero 
conductance according to (\ref{gkang}). Note that at $J=t'=0$, we obtain 
the opposite result. 

In the sequel, we want to recover these results from the continuum limit 
analysis of both models  in the Kondo regime.
In order to compare both models, it is instructive to write down explicitly 
the 
corresponding  Kondo models. In this regime, the tight-binding Anderson model 
(\ref{Hand})
reads: 
\begin{equation} 
   H_1=-t\sum_{j\leq -2}(c^\dagger_jc_{j+1}+h.c.)+-t\sum_{j\geq 1} 
   (c^\dagger_jc_{j+1}+h.c.) +J(c^\dagger_{-1}+c^\dagger_1) 
   {\vec \sigma \over 2}(c_{-1}+c_1)\cdot \vec S.\label{Hkondo}
\end{equation} 
Here the Kondo coupling is: 
\begin{equation} 
J=2t'^2\left[{1\over -\epsilon_d}+{1\over U+\epsilon_d}\right]. 
\end{equation} 
In general a potential scattering term is also induced, except 
in the case of the symmetric Anderson model, $U=-2\epsilon_d$. 
However, provided it is small, it has little effect on the 
transmission.

For the SCQD in the Kondo regime, the tight binding Anderson model (\ref{and}) simply reduces to
to the standard Kondo model: 
\beq
H_2=-t\sum\limits_{-\infty}^{+\infty} (c_{i+1}^\dag c_{i} +h.c.) + J~c_0^\dag { \vec{\s}\over 2} 
c_0\cdot\vec{S}~. 
\label{hk}
\eeq

At first sight, the two Kondo Hamiltonians (\ref{Hkondo}) and (\ref{hk}) 
just differ  in slight details. Nevertheless, when the Kondo coupling constant is switched off ($J=0$), the first model reduces to two independent
quantum wires and has therefore a  transmittance $T=0$
whereas the second one has a transmittance $T=1$.
Let us analyze the two models in the continuum limit.
We can first linearize the spectrum around the  Fermi points $\pm k_F$
and introduce left $(L)$ and right $(R)$ moving chiral fields:
\beq
\psi(x)\sim e^{-ik_Fx}\psi_{R}(x)+e^{ik_Fx}\psi_{L}(x)~,
\eeq
$\psi$ being the fermionic field associated with the operator $c$ in the continuum limit. The next step is to introduce the even/odd basis defined as follow:
\beq
\psi_{e/o}(x)={1\over \sqrt{2}}[\psi_L(x)\pm\psi_R(-x)].\label{psie}
\eeq 
Observe that these  two fields are left movers.
It may be advantageous in some situations
 to formulate the Kondo Hamiltonians as boundary problems.
 
For this purpose, we can fold the system by setting for $x>0$:
\bea
\psi_{L,e/o}(x)&=&\psi_{e/o}(x),\nn\\
\psi_{R,e/o}(x)&=&\psi_{e/o}(-x).\label{psile}
\eea
 All the fields are now defined on the  infinite half line $x>0$.
In both models, the Hamiltonians (\ref{Hkondo}) and (\ref{hk})
take the form:
\beq
H=H_0^{\rm odd}+H_0^{\rm even}+H_{\rm int}^{\rm even},\label{cont}
\eeq
with
\beq
H_0^{\rm even/odd}={v_F\over 2\pi}\int\limits_0^\infty dx
\left(\psi_{L,e/o}^\dag(x)i\partial_x\psi_{L,e/o}(x)
-(\psi_{R,e/o}^\dag(x)i\partial_x\psi_{R,e/o}(x)\right)~,
\label{continuum}
\eeq
the  Hamiltonians of free fermions. Notice that the Kondo interactions
couple only to the even sector which is the main advantage of this 
even/odd basis.
In order to derive the continuum limit for the interacting parts  of both
Hamiltonians, we need to pay attention to the boundary conditions at the 
origin before including the interactions. 
For the EQD, we had free boundary conditions on site $1$ and $-1$ on the lattice implying therefore $c_0=0$.
Consequently, we have $\psi(0^+)=\psi(0^-)=0$\footnote{we have used the notations $0^+$ and $0^-$ to differentiate fermions to the right (+) or left (-) of the dot} and therefore $\psi_R(0^+)=-\psi_L(0^+)$ and $\psi_R(0^-)=-\psi_L(0^-)$.
Using (\ref{psile}) and (\ref{psie}), we can infer that $\psi_{R,e}(0)=-\psi_{L,e}(0)$. Similarly, for the SCQD, we can prove that $\psi_{R,e}(0)=\psi_{L,e}(0)$ using  the continuity of $\psi_{L/R}(x)$ at $x=0$.\cite{Affleck} 

The boundary conditions can be summarized as follow:
\bea
{\rm EQD:}~~~ \psi_{R,e}(0)&=&-\psi_{L,e}(0), \label{bceqd}\\
{\rm SCQD:}~~~ \psi_{R,e}(0)&=&~~\psi_{L,e}(0)\label{bcsbqd}.
\eea
The interacting part of the EQD reads:
\bea
H_{int}^{even}&=&{J\over 2\pi}(\psi^\dag(1)+\psi^\dag(-1)){\vec \s\over 2}
\cdot\vec{S}
(\psi(1)+\psi(-1))\nn\\
&=&{2J\over 2\pi}(\psi^\dag_e(1)e^{-ik_F}+\psi^\dag_e(-1)e^{ik_F}){\vec \s\over 2}\cdot\vec{S} (\psi_e(1)e^{ik_F}+\psi_e(-1)e^{-ik_F}).
\eea
We then use the approximations $\psi_e(1)=\psi_{L,e}(1)\app \psi_{L,e}(0)$
and $\psi_e(-1)=\psi_{R,e}(1)\app \psi_{R,e}(0)$ and the boundary condition
(\ref{bceqd}).  
Finally, after these manipulations,
the associated interacting parts of the Kondo Hamiltonians read:
\beq
H_{int,i}^{even}=v_F\lambda_i\psi^{\dag}_{L,e}(0){\vec{\s}\over 2}\cdot\vec{S}\psi_{L,e}(0),\label{kinter}
\eeq
where the bare Kondo couplings are defined as follow:
\begin{equation} 
\lambda_1= {4J\sin^2k_F\over \pi v_F}={2J \sin k_F\over \pi t},
\label{lambda1def} 
\end{equation} 
\begin{equation} 
\lambda_2= {J\over \pi v_F}={J \over 2\pi t\sin k_F}.\label{lambda2def} 
\end{equation} 
Note that the derivation of (\ref{lambda2def}) comes simply from 
$\psi_{L,e}(0)=\psi(0)/\sqrt{2}$ for the SCQD. 
Therefore, both Kondo Hamiltonians have almost the same continuum limit, the main
difference being in the boundary conditions at the origin.
 
In both models the effective Kondo coupling renormalizes as: 
\begin{equation} 
d\lambda_i /d\ln D = -\lambda_i^2 + O(\lambda_i^3)~~~~i=1,2.\label{renorm}
\end{equation}
$D$ is the momentum space cutoff and effectively the bandwidth.
In both models, we expect the effective dimensionless kondo couplings
to renormalize to strong coupling namely $\lam_i\to \infty$.
On the lattice, this corresponds to an electron near the origin forming a singlet with
the impurity spin. The remaining electrons can propagate freely except that 
they cannot go into the same local orbital as the screening electron
since this would break the singlet and cost a large energy of
order $J$.  For the Hamiltonian $H_2$, the screening electron 
resides at site $0$.  Thus, at $J\to \infty$, the other 
electrons cannot cross the origin; they are trapped on either the 
positive or negative axis, corresponding to perfect reflectance. 
Formally, the low energy effective Hamiltonian is simply: 
\begin{equation} 
H_{2,low} =-t\sum_{j\leq -2}(c^\dagger_jc_{j+1}+h.c.) -t\sum_{j 
\geq 1 }(c^\dagger_jc_{j+1}+h.c.). 
\end{equation} 
Note that this effective Hamiltonian is similar to the embedded Kondo 
Hamiltonian (\ref{Hkondo}) at $J=0$.
On the other hand, for the Hamiltonian $H_1$, the screening electron 
is in a symmetric orbital on sites $1$ and $-1$
($ (c^1_s)^\dag=(c^\dagger_1+c^\dagger_{-1})/\sqrt{2}$).
 The 
remaining electrons are now allowed to occupy sites $1$ and $-1$ 
provided that they only go into the anti-symmetric state: 
 $c_a^\dag=(c_1^o)^\dag =(c^\dagger_1-c^\dagger_{-1})/\sqrt{2}$.
To obtain the low energy effective Hamiltonian in this case we must 
project the operators $c_{\pm 1}$ into the anti-symmetric state: 
\begin{equation} 
Pc_{\pm 1}P = \pm {c_a\over \sqrt{2}}.\end{equation} The effective 
low energy Hamiltonian then becomes: 
\begin{equation} 
H_{1,low}=-t\sum_{j\leq -3}(c^\dagger_jc_{j+1}+h.c.)-t\sum_{j\geq 
2}(c^\dagger_jc_{j+1}+h.c.)-{t\over \sqrt{2}}(-c^\dagger_{-2}c_a 
+c_a^\dagger c_2+h.c.).\label{Hlow}\end{equation} 
A simple calculation  
shows (see Appendix B)  that  
this effective scattering Hamiltonian has a transmission 
probability: 
\begin{equation} 
T(k)=\sin^2k.\label{trans}\end{equation} 
Thus when $k_F=\pi /2$, there is 
perfect transmission at the Fermi energy.  Since the conductance 
is determined by $T(k_F)$, this model exhibits perfect conductance 
at half-filling.  This corresponds to a transmission resonance for 
non-interacting electrons; it occurs at half-filling since the 
model has particle-hole symmetry in that case. 
 A major difference between the strong coupling limits of $H_1$ and 
$H_2$ is the sensitivity to particle-hole symmetry breaking of $H_1$ 
but not of $H_2$.  While $H_2$ exhibits  zero conductance for any 
filling factor, $H_1$ exhibits perfect conductance only at 
half-filling. 
In the limit of small bare Kondo coupling, the case  relevant to 
experiments, particle-hole symmetry breaking effects 
are small in both Kondo models, of order $J/D$. Similar terms are indeed 
neglected in both formula (\ref{glee}) and (\ref{gkang}) when approximating
$\de$ by ${\pi\over 2}\la n_d\ra$.  We would like
to emphasize that due to the fact that particle-hole 
symmetry breaking is an exactly marginal perturbation in 
renormalization group language, we can expect the conductance to 
be close to 1 for the Kondo Hamiltonian, $H_1$, at weak coupling and 
any filling factor. The special role played by particle-hole symmetry is
not apparent in the approximation leading to  eqs. (\ref{glee}) and (\ref{gkang}).

In our analysis, we have also supposed free electrons in the wire. 
The situation changes considerably if we include Coulomb interactions and has been extensively studied in ref. [\onlinecite{Kane}].
In this case, particle-hole symmetry breaking becomes  relevant  for repulsive 
Coulomb interactions.\cite{Kane,Gogolin} Nevertheless, the Kondo resonance can still be reached
(and therefore the unitary limit in the embedded dot) but only
for one special value of the gate voltage $V_g^*$ (and therefore $\eps_d^*$),
which would make the observation of the Kondo resonance a pretty 
delicate task. We will come back to this point latter in section \ref{interact}. 
The fact that the unitary limit has been successfully reached 
in 
[\onlinecite{Wiel}] may be an indication that Coulomb interaction effects
are very small in the leads. 

\begin{figure}[h]
\hspace* {2cm} \psfig{figure=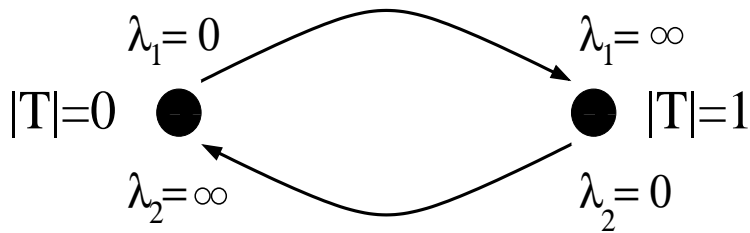,height=3cm,width=10cm}
\caption{Schematic flow for both  models between the $|T|=0$ fixed point
and the $|T|=1$ fixed point.}
\label{fixpt}
\end{figure}

A similar analysis of the strong coupling fixed point can be reproduced  using the field theory language. The strong coupling regime $\lam_i\to \infty$
corresponds to a $\pi/2$ phase shift in the even channel.
In terms of right and left movers, it changes the boundary conditions at
 the origin as follow:
\bea
&&\psi_{L,e}(0)=-\psi_{R,e}(0), ~~~\lam_1=0~~,~~\lam_2\to +\infty,\\
&&\psi_{L,e}(0)=~~\psi_{R,e}(0), ~~~\lam_2=0~~,~~\lam_1\to +\infty.
\eea
Therefore, the boundary conditions in the even channels of both
models get completely interchanged as it is schematically depicted in figure
\ref{fixpt}.
 The strong coupling fixed point
of one model corresponds to the weak coupling fixed point of the other and 
{\it vice versa}. In this analysis,  we have neglected all marginal operators
which correspond to particle-hole symmetry breaking operators.
The transmission probability depends on the scattering phase shift of both even and odd sector as follow:
\beq
|T|=\cos^2(\varphi_e-\varphi_o)
\eeq
where $\varphi_o=0,~ \forall~ \lam_i$ and $\varphi_e$ varies from $\pi/2$ ($\lambda_1=0,\lam_2=\infty$) to $0$ ($\lambda_1=+\infty,\lam_2=0$).
From this analysis, we recover the results obtained from   
the explicit analysis of the low energy effective lattice Hamiltonian
at strong coupling.

\section{Persistent currents for the embedded quantum dot} \label{pceqd}

We consider in this section the persistent current in the transmission Kondo 
model (\ref{Hkondo}).
 Thus we consider a closed ring of $l$ sites (including the 
impurity site, $0$) and apply a magnetic flux: 
\begin{equation} 
\Phi = (c/e)\alpha .\end{equation} (We work in units where $\hbar 
=1$. $e>0$ is the absolute value of the electron charge.) 
 In the Anderson model of Eq. (\ref{Hand}), this corresponds 
to modifying the hopping term between sites $j$ and $j+1$ by a 
phase factor $e^{i\alpha_j}$ such that $\sum_j\alpha_j=\alpha$. 
Which links carry the phase can always be changed by a gauge 
transformation, $c_j\to e^{i\phi_j}c_j$.  The total flux is, of 
course, gauge invariant. Choosing, for convenience, to put the 
phase factors on the 2 links connected to the impurity site, we 
obtain the Hamiltonian: 
\begin{equation} 
H=-t\sum_{j=1}^{l-2}(c^\dagger_jc_{j+1}+h.c.) 
-t'[c^\dagger_d(e^{-i\alpha /2}c_{l-1}+e^{i\alpha /2}c_1)+h.c.] 
   +\epsilon_dc^\dagger_dc_d 
   +Un_{d\uparrow}n_{d\downarrow}.\label{Handflux}\end{equation} 
   In the Kondo limit this becomes: 
\begin{equation} 
H=-t\sum_{j=1}^{l-2}(c^\dagger_jc_{j+1}+h.c.) +J(e^{i\alpha 
/2}c^\dagger_{l-1}+e^{-i\alpha /2}c^\dagger_1) 
   {\vec \sigma \over 2}(e^{-i\alpha /2}c_{l-1}+ 
   e^{i\alpha /2}c_1)\cdot \vec S.\label{Hkondoflux}\end{equation} 
   The zero temperature persistent current is determined 
from the flux dependence 
    of the ground state energy: 
    \begin{equation} 
    j=-c{dE_0\over d\Phi}=-e{dE_0\over d\alpha}. 
\label{current}\end{equation}

We expect that the persistent current is given by  universal
scaling functions of the variables $\alpha$ and $l/\xi_K$ in
the limit $l,\xi_K >>a$ (where $a$ is the lattice constant, which
we have set to 1):
\begin{equation}
lj= f(\alpha , \xi_K/l).\label{scale}\end{equation}
 These universal functions depend on the
parity of the electron number, $N$ and also, of course, on
whether we consider the EQD or the SCQD.  This behavior is
a rather immediate consequence of Eq. (\ref{current}), expressing the
persistent current as the derivative of the groundstate
energy with respect to the phase $\alpha$ and the
applicability of usual continuum field theory scaling
arguments.  Since $\alpha$
is a phase, associated with the boundary conditions, we
expect it to have zero anomalous dimension.  The universal
part of the ground state energy is expected to have the
canonical dimension of 1/(length), and thus the result
follows from standard finite size scaling hypotheses.
   $lj$ may be expressed as a universal function
of $\alpha$ and either the dimensionless renormalized 
Kondo coupling $\lambda_{eff}(l)$ or equivalently
of the ratio $\xi_K/l$.  A non-zero value of $a/l$ leads to
the appearance of various irrelevant operators in the effective
Hamiltonian which produce corrections to $lj$, down by
powers of $a/l$.  This scaling form implies that $j$
can be calculated using renormalization group improved
perturbation theory, at large $\xi_K/l$ where
$\lambda_{eff}(l) <<1$.  It also implies that the current
at small $\xi_L/l$ will be a universal characteristic
of the strong coupling RG fixed point. 

We now want to calculate the 
persistent current in perturbation theory in $J$.
When $J=0$ the impurity spin is decoupled from the rest of the chain 
which effectively consists of $N-1$ electrons on 
$l-1$ sites with open boundary conditions. 
The ground state at 
$0^{th}$ order in $J$, consists of a product of the impurity 
state (which may be spin up or spin down) and the filled 
Fermi sea for the rest of the system.  The $0^{th}$ order ground state 
energy is independent of flux.  The single particle 
energy levels have energy $E_n=-2t\cos k_n$, with $k_n=\pi 
n/l$, $n=1,2,3,\ldots,l-1$.  
The annihilation 
operator appearing in the interaction may be expanded in Fourier modes: 
\begin{equation} 
\chi \equiv (e^{i\alpha /2}c_1+e^{-i\alpha /2}c_{l-1}) 
=\sqrt{2\over l}\sum_k[e^{i\alpha /2}\sin k + e^{-i\alpha 
/2}\sin k(l-1)] c_k.\label{defchi}
\end{equation}

From this stage, we have to distinguish
the cases $N$ even and $N$ odd. Here $N$ is the total number of electrons, 
{\it including the 
impurity spin.}
Let us first consider the case of an even number of electrons.

\subsection{Perturbation theory for N even}
For even $N$, the levels are 
doubly occupied for $n\leq N/2-1$.  However, the state with 
$k=\pi N/2$ contains only one electron which may have spin up 
or down.  Thus the ground state is 4-fold degenerate at $J=0$. 
This degeneracy is split in first order in $J$.
It is convenient to decompose 
$\chi$ into the term involving $k=k_F=\pi N/2l$, which we label 
$\chi_0$, plus all other terms, which we label $\chi{'}$. 
\begin{equation} 
\chi_0 = \sqrt{2\over l} \sin k_F\left[ e^{i\alpha /2}-(-1)^{N/2} 
e^{-i\alpha /2}\right ]c_{k_F}.\end{equation} 
Only $\chi_0$ contributes in first 
order perturbation theory.  Keeping only this contribution, 
the Kondo interaction in Eq. (\ref{Hkondoflux}) reduces to: 
\begin{equation} 
H_{int} = {8J\over l}\sin^2(k_F) \sin^2(\tilde \alpha /2) 
c^\dagger_{k_F}{\vec \sigma \over 2}c_{k_F}\cdot \vec S, 
\label{HintkF}
\end{equation} 
 where 
\begin{equation} 
\tilde \alpha \equiv \alpha + k_Fl=\alpha + \pi 
N/2,\label{tilde}
\end{equation} 
and hence: 
    \begin{equation} 
    \sin^2\tilde \alpha /2 = \left[ 1-(-1)^{N/2}\cos \alpha 
    \right] /2.\end{equation} 

Diagonalizing $H_{int}$ in the degenerate subspace picks out 
the singlet state, with the first order ground state energy: 
\begin{equation} 
E_0=\hbox{constant} -{6J\over l}\sin^2k_F\sin^2\tilde \alpha /2. 
\end{equation} 
Thus the persistent current, to $O(J)$ is given by: 
\begin{equation} 
j_e(\alpha ) ={3eJ\over l}\sin^2k_F \sin \tilde \alpha . 
\end{equation} 
We see from Eq. (\ref{tilde}) that, at large $l$, the current 
oscillates in $N$ with period $4$ and that the current for 
$N/2$ odd is given by the current for $N/2$ even with a shift 
of $\alpha$ by $\pi$: 
\begin{equation} 
j_{e,N+2}(\alpha ) \app j_{e,N}(\alpha + \pi 
).\label{pishift}
\end{equation}
Notice that Eq. (\ref{pishift}) is indeed exact 
to all order of the perturbative theory ignoring corrections of $O(a/l)$.  Indeed, we can show that it is a property
of the two different types of propagators involved in the calculation 
of the ground state at higher order of perturbation theory (see eqs. 
(\ref{G(tau)}) and ({\ref{propa2})), therefore a property of the ground state energy itself, and thus a property of the persistent current.

To proceed to higher orders of perturbation in $J$, 
it is convenient to  include 
the $\chi_0$ part of the Kondo interaction, in Eq. (\ref{HintkF}) 
in the unperturbed Hamiltonian, $H_0$. This Hamiltonian can be 
diagonalized exactly.  The ground state has all levels doubly 
occupied or else empty except for the $k_F$ level. It has one 
electron in the $k_F$-level which forms a singlet with the 
impurity spin: 
\begin{equation} 
|s> = (a_{k_F,\uparrow}^\dagger |\Downarrow >- 
a_{k_F,\downarrow}^\dagger |\Uparrow >)/\sqrt{2}.\end{equation} 
(Here the double arrows label the spin state of the impurity.) 
The remaining perturbation, 
$H_{int}$, has a term quadratic in $\chi {'}$ and cross-terms 
containing one factor of $\chi_0$ and one factor of $\chi {'}$. 
 The second order correction to the ground state 
energy can be written in the standard form: 
\begin{equation} 
E_0^{(2)}=-{1\over 2!}\int_{-\infty}^\infty 
 d\tau {\cal T}<s|H_{int}(\tau )H_{int}(0)|s>. 
\label{e02}
\end{equation} 
The propagator for the field $\chi {'}$ is, for $\tau >0$: 
\begin{equation} 
 <\chi {'}^\dagger_{\alpha} (\tau )\chi {'}_{\beta}(0)> 
 \equiv \delta_{\ga \beta}  G(\tau ) 
   ={4\delta_{\ga \beta} 
\over l}\sum_{n=1}^{N/2-1}[1-(-1)^n\cos \alpha ]\sin^2(\pi n/l) 
e^{\xi_n\tau }, \label{propa1}
\end{equation} 
where 
\begin{equation} 
\xi_n\equiv -2t\cos (\pi n/l)+2t\cos (\pi N/2l).\end{equation} 
At large $\tau$ the sum is dominated by terms near the Fermi surface, 
$n=N/2-m$, with $m<<N$, so we may write: 
\begin{eqnarray} 
G(\tau )&\approx& {4\over l}\sin^2k_F\sum_{m=1}^{l-1} 
[1-(-1)^{N/2-m} 
\cos \alpha ]e^{-v_F\tau \pi m/l}\nonumber \\ 
&=&{4\over l}\sin^2k_F\left[ {1\over e^{\pi v_F\tau /l}-1} + 
{\cos \tilde \alpha \over 
 e^{\pi v_F\tau /l}+1}\right],\ \  (\tau >0).\label{G(tau)} 
\end{eqnarray} 
It can be easily seen that: 
\begin{equation} 
{\cal T}<\chi {'}^\dagger (\tau )\chi {'}(0)>= 
{\cal T}<\chi {'} (\tau )\chi {'}^\dagger (0)>=G(\tau )= 
\epsilon (\tau )G(|\tau |),\label{Gneg} 
\end{equation} 
where $G(|\tau |)$ is given approximately by Eq. (\ref{G(tau)}),
and $\eps(\tau)=1$ if $\tau>0$ and $\eps(\tau)=-1$ if $\tau<0$.

We have calculated in the appendix \ref{eqde} $j_e(\alpha )$ using (\ref{e02}) and the  propagator (\ref{G(tau)}) 
 to second order in $J$, for $N$ even and large $l$. We have ignored corrections of order $O(a/l)$, with $a$ the lattice spacing.  
The result reads: 
\beq 
j_e(\alpha ) = {3\pi v_F e\over 4  l}\left[\sin \tilde \alpha 
[\lambda + \lambda^2\ln (lc)] +(1/4+\ln 2)\lambda^2\sin 
2\tilde \alpha \right] + O(\lambda^3)\ \  (N \ \hbox{even})\label{jperte}
\eeq
 where $c$ is a 
constant of O(1) which we have not determined.

Let us now analyze the case $N$ odd.

\subsection{Perturbation theory for N odd}
In this case there is an exact two fold degeneracy of the 
ground state, which has $S=1/2$.  The unperturbed ($J=0$) 
ground state has all free electron levels doubly occupied for 
$k<k_F\equiv \pi N/2$ and empty for $k>k_F$.  (We choose $k_F$ to 
lie exactly half-way between the highest filled and lowest empty 
levels.) Now first order perturbation theory vanishes since it 
necessarily gives states with a particle-hole excitation.  Second 
order perturbation theory is now straightforward since the 
ground state is unique, once we specify a value of 
$S^z_{total}=1/2$ for example. The exact time-ordered Green's 
function is now, for $\tau >0$: 
\begin{equation} 
{\cal T} <\chi^\dagger_{\alpha} (\tau )\chi_{\beta}(0)> 
 \equiv \delta_{\alpha \beta}  G(\tau ) 
   ={4\delta_{\alpha \beta} 
\over l}\sum_{n=1}^{(N-1)/2}[1-(-1)^n\cos \alpha ]\sin^2(\pi n/l) 
e^{\xi_n\tau },\ \  (\tau >0).\end{equation} 
where: 
\begin{equation} 
\xi_n\equiv -2t \cos \pi n/l +2t \cos \pi N/2l.\end{equation} 
Again, at large $\tau$ this sum is dominated by states near the 
Fermi surface so we can approximate: 
\begin{eqnarray} 
G(\tau )&\approx& {4\over l}\sin^2k_F\sum_{m=0}^{l-1}
[1-(-1)^{(N-1)/2-m} 
\cos \alpha ]e^{-v_F\tau \pi (m+1/2)/l}\nonumber \\ 
&=&{2\over l}\sin^2k_F\left[ {1\over \sinh (\pi v_F \tau /2l)} - 
{(-1)^{(N-1)/2}\cos  \alpha \over \cosh (\pi v_F \tau /2l)}\right], 
\ \  (\tau >0).\label{G(tauodd)} \end{eqnarray} 
For $\tau <0$, we sum over the unoccupied levels, with 
$n=(N+1)/2+m$, $m=0,1,2,\ldots$.  In this case: 
\begin{equation} 
{\cal T} <\chi^\dagger_{\alpha} (\tau )\chi_{\beta}(0)> 
 \equiv \delta_{\alpha \beta}  G(\tau ) 
   =-{4\delta_{\alpha \beta} 
\over l}\sum_{n=(N+1)/2}^{l-1}[1-(-1)^n\cos \alpha ]\sin^2(\pi 
n/l) e^{\xi_n\tau },\ \  (\tau <0).\end{equation} At large $|\tau 
| $, this becomes: 
\begin{eqnarray} 
G(\tau )&\approx& -{4\over l}\sin^2k_F\sum_{m=0}^\infty 
[1-(-1)^{(N+1)/2+m} 
\cos \alpha ]e^{v_F\tau \pi (m+1/2)/l}\nonumber \\ 
&=&{2\over l}\sin^2k_F\left[ {1\over \sinh (\pi v_F \tau /2l)} - 
{(-1)^{(N-1)/2}\cos  \alpha \over \cosh (\pi v_F \tau /2l)}\right], 
\ \  (\tau <0).\label{G(tauoddneg)} \end{eqnarray} 
Thus $G(\tau )$ is given by the same approximate expression, 
Eq. (\ref{G(tauodd)}), for either sign of $\tau$. 
As usual, we may write: 
\begin{equation} 
{\cal T} <\chi_{\alpha}(\tau )\chi^\dagger_{\beta}(0)> = - 
\delta_{\alpha \beta}G(-\tau ).\end{equation} 
The second order term in the ground state energy is: 
\begin{equation} 
E_0^{(2)} = {-J^2\over 2!} \sum_{a,b} 
\hbox{tr}\left( {\sigma^a\over 2}{\sigma^b\over 2}\right) <S^aS^b> 
\int_{-\infty}^\infty d\tau G(\tau )[-G(-\tau )].\end{equation} 
If we use our large $\tau$ expression, Eq. (\ref{G(tauodd)}) for 
$G(\tau )$, then we get an $\alpha$-independent term whose integral 
diverges at $\tau =0$.  This is just an artifact of the large-$\tau$ 
approximation to $G(\tau )$ and in any event, this term doesn't 
contribute to the persistent current.  The other term is 
$\alpha$-dependent and finite: 
\begin{equation} 
E_0^{(2)} = \hbox{constant} +\cos^2\alpha {3J^2\sin^4k_F\over 
4l^2} \int_{-\infty}^\infty d\tau {1\over \cosh^2(\pi v_F \tau 
/2l)} =\hbox{constant} + \cos^2\alpha {3J^2\sin^4 k_F\over 
lv_F\pi}.\label{E02odd}\end{equation} 
Note that, in this case, the 
result does not depend on the parity of $(N-1)/2$ and is periodic 
in $\alpha$ with period $\pi$. We have generalized this result to third order
in $J$ in appendix \ref{eqdo}. Unfortunately, we have not been able to
prove it to all orders of the perturbation theory.
From the ground state energy, it is then straightforward to infer the
persistent current:
\beq
j_o(\alpha )={3\pi v_F e\over 16  l}\sin 2\alpha [\lambda^2+2\lambda^3\ln 
(lc')]+O(\lambda^4)\ \  (N\ \hbox{odd}),\label{jperto} 
\eeq
where $c'$ is another undetermined constant.

Several remarks about the persistent currents expressions
(\ref{jperte}) and (\ref{jperto}) are in order.  First of 
  all, we note a large parity effect- at small $\lambda$ the 
  persistent current is much larger for even $N$ than for odd $N$. 
  Furthermore, as already noticed the periodicity is different in both cases.

We have chosen to present the calculations using the lattice propagator
and then consider the limit $l\gg a$, $a$ the lattice spacing.
We have also checked that we obtain the same results using directly 
the continuum limit representation of the Kondo Hamiltonian (\ref{Hkondoflux}).
The best strategy is first to unfold the chain with free boundary conditions
in sites $1$ and $l-1$ in order to work with left movers only (the size of the chain 
becomes  so far $2l$).
In this representation, the linearized Hamiltonian (\ref{Hkondoflux}) takes the form: 
\bea
H&=&{v_F\over 2\pi}\int\limits_{-l}^{l}dx~ \psi_L^\dagger i\partial_x\psi_l\nn\\
&+&v_F{\lambda_1\over 2} \left[\psi_L^\dag(0)e^{-i(\al-k_Fl)/2}-\psi_L^\dag(l) e^{i(\al-k_Fl)/2}
\right]{\vec{\s}\over 2}\cdot \vec{S}\left[\psi_L(0)e^{i(\al-k_Fl)/2}-\psi_L(l) e^{-i(\al-k_Fl)/2}
\right].
\eea
From this point, the perturbative calculations become  very similar to what was presented in the text and similar results are obtained using the propagators
involving left movers.

\subsection{Renormalization group arguments and the strong coupling limit}

In this section, we want to comment  on the $l$-dependence
of the persistent currents (\ref{jperto}) and (\ref{jperte}). 
In both cases, to the order in $\lambda$ that we have worked, the 
result has the form: 
\begin{equation} 
j_{e/o}= {ev_F\over l} f_{e/o}[\lambda_{eff}(l),\tilde \alpha ],
\label{univers} 
 \end{equation} 
where 
$\lambda_{eff}(l)$ is the renormalized coupling constant at scale 
$l$: 
\begin{equation} 
\lambda_{eff}(l) =\lambda + \lambda^2 \ln l + \ldots 
\end{equation}  
If this result persists to all orders in 
perturbation theory, this would imply the scaling form for $j$ 
we have anticipated on general arguments in Eq. (\ref{scale}) 
(which is valid in 
the limit of small bare coupling $\lambda$).  
In particular, this implies that, at 
$l<<\xi_K$, the perturbative result is valid.  The finite size of 
the ring cuts off the infrared divergences of perturbation theory. 
On the other hand, when $l$ is of order of $\xi_K$ or greater, 
perturbation theory breaks down.  It is well-known from numerical 
renormalization group, Bethe ansatz and other calculations that 
one may think of the effective coupling constant as renormalizing 
to $\infty$ in the low energy effective Hamiltonian. Technically, 
this corresponds to a fixed point of the boundary renormalization 
group. Thus it is very reasonable to expect that the persistent 
current for $l>>\xi_K$ can be obtained by simply taking the limit 
$J\to \infty$.  This is expected to be valid because the 
persistent current is determined only by the low energy properties 
of the Hamiltonian and these are correctly represented by the 
infinite coupling fixed point. 
 
It is a straightforward matter to calculate the persistent current 
for the tight-binding Hamiltonian of Eq. (\ref{Hkondo}) at 
infinite $J$. In this limit we may simply use the low energy 
effective Hamiltonian of Eq. (\ref{Hlow}). It is convenient to 
think of this as being a simple free electron model defined on 
$(l-2)$ sites labeled $a, 2,3,\ldots ,l-2$.  (The impurity site 
is eliminated and the 2 neighboring sites, $1$ and $l-1$ have 
effectively collapsed to one site due to the projection onto the 
odd parity linear combination.) We can add a flux to the model by 
changing the phase of any hopping term. In particular, since the 
hopping term between sites $l-2$ and $a$ already has a reversed 
sign, it is convenient to change the phase of that term, resulting 
in the model: 
\begin{equation} 
H_{low}=-t\sum_{j\leq -3}(c^\dagger_jc_{j+1}+h.c.)-t\sum_{j\geq 
2}(c^\dagger_jc_{j+1}+h.c.)-{t\over \sqrt{2}}(e^{i(\alpha + \pi ) 
}c^\dagger_{-2}c_a +c^\dagger_ac_2+h.c.).\label{Hlowalpha} 
\end{equation} 
We see that this is a standard potential scattering 
model with an effective flux of $\alpha +\pi$.  The $\pi$ flux 
shift is a consequence of the projection onto the odd linear 
combination. As shown by Gogolin and Prokof'ev \cite{Gogolin} the persistent 
current at zero temperature and large $l$ is completely determined 
by $T(k_F)$, the transmission probability at the Fermi surface,
in such a potential scattering model (defined at $\tilde \al=0$). 
From Eq. (\ref{trans}), at half-filling, $k_F=\pi /2$, 
$T(k_F)=1$.  For this value of $T(k_F)$, the persistent current is 
the same as for an ideal periodic ring. 
 By summing over the energies of the filled Fermi 
sea,with momenta $k=(2\pi n+\alpha )/l$, $n=0,\pm 1,\pm 2,\ldots$, 
it can be shown that this gives, for $M$ spinless fermions, an 
energy: 
\begin{eqnarray} 
E_0 &=& -2t\cos([\al]/l){\sin2\pi(M+{1\over 2})/l\over \sin\pi/l}=
\hbox{constant} + {v_F\over 2\pi l}[\alpha ]^2\ \  (M\ 
\hbox{odd})\label{freeenodd} \\ 
&=& -2t\cos([\al-\pi]/l){\sin2\pi(M)/l\over \sin\pi/l}=
\hbox{constant} + {v_F\over 2\pi l}[\alpha -\pi ]^2\ \  (M\ 
\hbox{even}),\label{freeeneven}\end{eqnarray} where $[\theta ]$ 
equals the principal part of the angle $\theta$, which lies 
between $-\pi$ and $\pi$ and jumps by $-2\pi$ at $\theta = 
(2n+1)\pi$.  ie. 
\begin{eqnarray} 
[\theta ] &=& \theta ,\ \  (|\theta | <\pi )\nonumber \\ &=& 
\theta -2\pi ,  \ \  (\pi <\theta <3\pi ),\end{eqnarray} etc. Thus 
$E_0(\alpha )$ has minima at $\alpha = 2\pi n$, for $M$ odd and at 
$2\pi (n+1/2)$ for $M$ even.  When we include electron spin, for 
non-interacting electrons, we find a 4-fold periodicity in the 
{\it total} number of electrons, $N$. If $N$ is odd, then the 
numbers of electrons of opposite spin are necessarily of opposite 
parity, so the energy is the sum of the two terms in Eq. 
(\ref{freeenodd}) and (\ref{freeeneven}).  On the other hand, if 
$N$ is even, then the numbers of electrons of spin up and down are 
equal in the ground state, so the energy is  twice Eq. 
(\ref{freeenodd}) if $N/2$ is odd and twice Eq. (\ref{freeeneven}) 
if $N/2$ is even. 
 
To apply this free electron result to the large $l$ (i.e. strong 
coupling) limit of the Kondo model, we should take into account the 
shift, $\alpha\to \alpha +\pi$ in Eq. (\ref{Hlowalpha}), and also 
the fact that two electrons are removed from the low energy 
subspace corresponding to the impurity spin and the screening 
electron.   These two effects actually cancel. 
Writing the result in terms of $\tilde \alpha $ 
defined in Eq. (\ref{tilde}), we thus obtain the result for  $l>>\xi_K$: 
\begin{eqnarray} 
j_e(\alpha ) &=&-{2ev_F\over \pi l}[\tilde \alpha  -\pi ],\ \  (N \ 
\hbox{even})\nonumber \\ 
j_o(\alpha ) &=& -{ev_F\over \pi l}([\alpha ] + [\alpha - \pi ])\ \ 
(N \ \hbox{odd}) \label{pcfree}. 
\end{eqnarray}  

The current is paramagnetic (i.e. positive for small positive $\Phi$, 
corresponding to a current which {\it increases} the flux through the ring) 
for $N$ odd or for $N$ and $N/2$ even.  It is diamagnetic for 
$N$ even and $N/2$ odd.  This is exactly the same situation which we found at 
$l<<\xi_K$ in eqs: (\ref{jperte}) and (\ref{jperto}).  It seems plausible that this feature 
is independent of $\xi_K/l$. We have plotted the persistent currents for
$N/2$ even and for $N$ odd in Figures \ref{empce} and \ref{empco}.
 
\begin{figure}[h]
\hspace{3cm}\psfig{figure=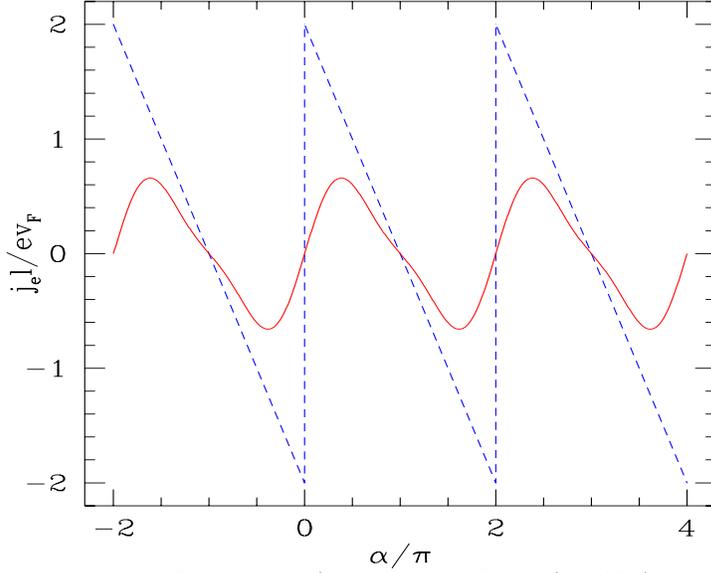,height=8cm,width=10cm,angle=0}
\caption{ Persistent current of the EQD versus $\al/\pi$ for  $N=4p$  
 for $\xi_K/l\app 50$ (solid line) and for $\xi_K/l\ll 1$ (dashed line).
The persistent current for $N=4p+2$ is obtained by a simple translation of 
$\pi$. Note that $\lam_{eff}(l)=1/\ln(\xi_K/l)$.}
\label{empce}
\end{figure}

\begin{figure}[h]
\hspace{3cm}\psfig{figure=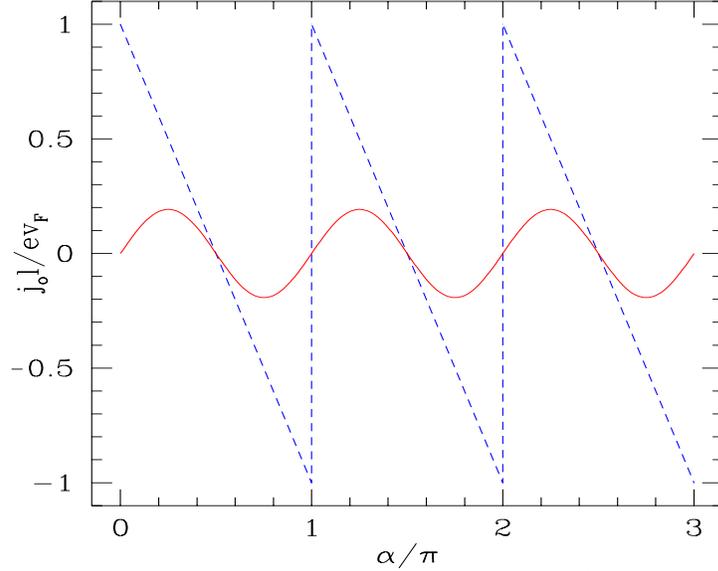,height=8cm,width=10cm,angle=0}
\caption{ Persistent current of the EQD versus $\al/\pi$ for  N odd and for
 for $\xi_K/l\app 50$ (solid line) and for $\xi_K/l\ll 1$ (dashed line). 
$j_o$  has been multiplied by $\times 5$  for
visibility.}
\label{empco}
\end{figure}

\subsection{Particle-hole symmetry breaking}\label{phb}

A crucial aspect of Kondo physics is the {\it  exact marginality} 
of particle-hole (P-H) symmetry breaking. By exact marginality 
we mean that the P-H symmetry breaking coupling constants 
grow neither larger nor smaller under renormalization to 
all orders in perturbation theory.  (By contrast the Kondo 
coupling itself is marginally relevant, since the 
corresponding RG $\beta$-function vanishes in first order but
is non-zero in second order.)

 The model is 
easiest to analyze when it has {\it exact} P-H symmetry: for instance 
the Anderson model of Eq. (2) with $\epsilon_D=-U/2$ 
{\it at half-filling}.  Then various properties of the low energy 
physics follow exactly from symmetry.  The average 
occupation number on every site $<n_i>$ has exactly the 
P-H symmetric value 1.  This implies that the formation 
of the Kondo screening cloud has no effect whatsoever on the charge 
density.  The peak in the single electron density of states 
(Kondo resonance) occurs {\it exactly} at the Fermi energy, $E=0$.  
The phase shift at the Fermi surface is {\it exactly} $\pi /2$, as 
follows, for instance, from Langreth's application of the 
Friedel sum rule.  Thus, the transmission probability at the 
Fermi surface is exactly 1 and the persistent current should have 
exactly the value for a perfect ring, Eq. (\ref{pcfree}) in the 
limit $\xi_K/l\to 0$.  

Once P-H symmetry is broken, all these properties change somewhat.  
However, the important result is that the changes to these 
properties are small, of order the dimensionless bare 
Kondo coupling. This is a non-trivial fact given that 
these are all low energy properties and that  the effective 
Kondo coupling at low energies is expected to renormalize 
to large values.   A convenient way of seeing this\cite{PHrefs,Barzykin}
 is from 
taking the continuum limit of the Kondo model and then 
applying bosonization (either Abelian or non-Abelian) which leads 
immediately to spin-charge separation.  One can then see that 
the Kondo interaction only involves the spin boson and the 
charge boson is apparently completely unaffected.  If we start 
with an asymmetric Anderson model, $\epsilon_d\neq -U/2$, then 
the resulting large-$U$ Kondo model contains a potential 
scattering term, of strength $V$, 
which is typically of the same order of magnitude 
as the Kondo coupling $J$.  Under bosonization, this becomes 
$\propto V\partial_x \phi_c$ where $\phi_c$ is the charge boson.  Since 
this is linear in $\phi_c$ it leaves the charge sector non-interacting 
and hence is strictly marginal.  By factorizing  the electron 
Green's function into charge and spin factors, it can be seen that 
this has the effect of changing the phase shift at the Fermi surface 
by an amount of order $V/D$. This is typically a small 
dimensionless number of order $t'^2/Ut\sim J/t\sim \lambda_0$.  
 There is a corresponding change 
in the electron density in the vicinity of the impurity.  This, in 
fact follows from the Friedel sum rule as pointed out by Langreth. 
The phase shift at the Fermi surface is related to 
the total charge displaced near the impurity.  In the Kondo 
limit $<n_d>$ is exactly 1 and the small change in the phase shift 
is associated with a small change in the charge density at the 
nearby sites, i.e. in the charge density of the conduction electrons, 
rather than the ``d-electrons''.  Even if $\epsilon_d=-U/2$ so that 
the Anderson model is nominally ``symmetric'', a P-H symmetry 
breaking electron density, different than 1, has a similar effect.  
In the Kondo limit, an effective potential scattering term is 
generated at second order in the bare coupling $\lambda$.
However, we again expect no large renormalization of this 
potential scattering term for the reason discussed above.  
In these cases the transmission probability at the Fermi energy, will 
be slightly less than 1 and the persistent current will be 
slightly modified from that of Eq. (\ref{pcfree}) even in the limit 
$\xi_K/l<<1$.  However, these modifications will be small, of 
order $V/D$.  Thus, as a function of $\epsilon_d$, for other 
parameters fixed, the transmission probability at zero temperature is 
expected to have a broad plateau where it is only slightly 
less than 1 for a range of $\epsilon_d$ of $O(U)$.  Changing 
the filling factor only slightly modifies the shape of this plateau.

On the other hand, things are very different if one considers 
P-H symmetry breaking in  
the Kondo limit with a bare Kondo coupling which is not small.  
In this case the potential scattering term $V/D$ becomes of 
order 1 and the P-H symmetry breaking effects are large.  This is true 
regardless of whether or not $\epsilon_d=-U/2$.  P-H symmetry 
breaking from the filling factor also has a large effect.    
Now the transmission probability is not, in general, close to 1 
even when $\xi_K/l\to 0$. The modification of the 
charge density on the sites near the impurity is large, although 
$\la n_d\ra$ itself remains at exactly 1 in the Kondo limit.  
 In the extreme limit where the 
Kondo coupling goes to $\infty$, it is easy to calculate the 
transmission probability, and hence the persistent current, 
exactly. 
Using Eq. (\ref{trans}) and Gogolin and Proko'fev exact result
\cite{Gogolin}, 
the persistent current at arbitrary filling becomes in this limit:
\bea
j_e(\al)&=&-{2ev_F\over \pi l}~
{\sqrt{T_F}\sin\al\over \sqrt{1-T_F\cos^2\al}}
\left(\arccos (\sqrt{T_F}\cos\al)-\pi\de_{N,4p}\right),~~(N~even),\\ 
j_o(\al)&=&-{2ev_F\over \pi l}
~{\sqrt{T_F}\sin\al\over \sqrt{1-T_F\cos^2\al}}
\left(2 \arccos (\sqrt{T_F}\cos\al)-\pi \right), ~~(N~odd) \label{gog},
\eea
with $T_F=T(k_F)=\sin^2(k_F)$.
When $k_F\ne \pi/2$, these expressions differ considerably  from (\ref{pcfree}).
\subsection{Inclusion of interactions in the ring }\label{interact}

Electron-electron  interactions in the ring have a dramatic effect
on the conductance of the dot, and hence on the persistent
current.\cite{Kane,Gogolin}  This can be seen by considering the
strong coupling effective Hamiltonian of Eq. (\ref{Hlow}) for the EQD
case.  Now consider the effect of P-H symmetry breaking.  For
instance, if we choose half-filling, then we could include P-H
symmetry breaking by a local potential scattering term at the
origin:
\begin{equation} 
H_{1,low}\to H_{1,low}+Vc_a^\dagger c_a.
\end{equation}
The corresponding continuum limit Hamiltonian is just the free Dirac
fermion Hamiltonian defined on $-\infty<x<+\infty$, 
 with a back scattering term,
$V[\psi^\dagger_L(0)\psi_R(0)+h.c.]$ added.  If $V$ is small, this
has little effect, reducing the conductance by an amount of
$O(V^2)$.  On the other hand, if we include electron-electron
interactions in the ring, the potential scattering has a much
larger effect.  We can simply use the analysis of Kane and Fisher \cite{Kane}
 (see also Ref. [\onlinecite{Wong}]).
Upon bosonization, and introduction of spin and charge bosons, we
parameterize the strength of the (screened, short range)
interactions by a parameter, $g_{\rho}$ related to the
compactification radius of the charge boson by $g_{\rho}=1/\pi
R_c^2$.  For zero interactions, $g_{\rho}=2$.  Repulsive
interactions decrease $g_\rho$.  (We assume that the interactions
preserve the full spin rotation symmetry so that the corresponding
parameter for the spin boson is unchanged from its non-interacting
value, $g_{\sigma}=1/\pi R_s^2=2$.)  The renormalization group
scaling dimension of this backscattering term is:
\begin{equation}
x=g_\rho /4+1/2.\label{dimback} 
\end{equation} 
This boundary
interaction is exactly marginal for zero bulk interactions,
relevant for repulsive interactions and irrelevant for attractive
interactions. Thus, with repulsive bulk interactions, this back
scattering parameter, $V$,  renormalizes to large values at low
energy scales corresponding to perfect reflection of spin and
charge at the quantum dot.  This implies that $jl\to 0$ for large
$l$.  The consistency of this assertion can be checked by
considering the stability of the perfectly reflecting fixed
point. \cite{Kane} At this fixed point, the ring is effectively
severed at the quantum dot so that open boundary conditions can be
applied to the fermion fields to the left and right of the dot.
The leading irrelevant operator at this fixed point corresponds to
a weak hopping process between the two ends of the severed chain,
on either side of the quantum dot.  This operator has scaling
dimension:
\begin{equation}
x=1/g_{\rho}+1/2,\label{xsevered}
\end{equation} 
which is $>1$ in the repulsive case.
Hence this hopping process is irrelevant, for repulsive
interactions, demonstrating the consistency of the assertion about
the RG flow.  We may think of this stable fixed point, where the
Kondo coupling and the potential scattering, have both
renormalized to $\infty$, as corresponding to an electron in the
even orbital at the original forming a spin singlet with the
quantum dot with the odd orbital either empty or doubly occupied
depending on the sign of $V$. From this RG analysis of the
perfectly reflecting fixed point we can deduce the scaling of the
persistent current with length.\cite{Gogolin}  The weak hopping
amplitude scales with length as
\begin{equation} t_{eff}(l)\propto l^{1/2-1/g_{\rho}}.\end{equation} The
conductance, being second order in $t_{eff}$, scales with length
as $l^{1-2/g_\rho }$.  On the other hand, $jl$ scales as
$t_{eff}$, going as the square root of the conductance, or the
transmission probability at the Fermi surface, $T_F$ \cite{Gogolin}
.  Thus, we
expect \cite{Gogolin}:
\begin{equation}
j\propto l^{-(1/g_\rho +1/2)}\sin \alpha .
\end{equation} 
This
scales to zero faster than $1/l$ for repulsive interactions. It is
also interesting to consider the behavior for small $V$, close to
the resonance.  It follows from Eq. (\ref{dimback}) that the
effective back-scattering potential at length $l$ is given by:
\begin{equation}
V_{eff}(l)\propto l^{1/2-g_\rho /4}.
\end{equation} 
Thus the
persistent current should depend on $V$ and $l$ as:
\begin{equation}
jl= f(Vl^{1/2-g_\rho /4},\alpha ).\end{equation} 
(This result only
holds in the case $\xi_K<<l$.)  In particular, this implies that
the width of the peak in $j$, as a function of $V$,
 scales as $l^{-1/2+g_\rho /4}$.
Thus, for weakly repulsive interactions, $g_{\rho}\leq 2$, $j$
will have a broad maximum as a function of gate voltage for
intermediate $l$  but this maximum will sharpen up with increasing
ring length.  The current at the maximum should be that of an
ideal ring with no quantum dot.

We note that, provided $g_{\rho}$ is not too small, and assuming
symmetric leads, there is only one relevant operator at the
resonant fixed point.  Thus, tuning one parameter, such as the
gate voltage, should be sufficient to pass through the resonance.
Strict P-H symmetry is not necessary to get perfect conductance;
even away from half-filling there should be one value of gate
voltage where perfect conductance occurs.  With asymmetric leads
$jl$ will scale to 0 with $l$ for all gate voltages.

Further insight can be obtained by considering the weak Kondo
coupling fixed point, with bulk interactions.  Now the Kondo
interaction breaks up into two parts which have different RG
scaling dimension: the term that reflects an electron coming
from the left or right and the term that transmits an electron.
The folding transformation that we mentioned in Sec. II is not
useful anymore because it would turn short range bulk interactions
into infinite range interactions. We use subscripts $+$ and $-$
to label the fermion fields to to the right or left of the dot.
 The open boundary conditions,
at zero Kondo coupling, imply:
\begin{equation}
\psi_{L\pm }(0)=-\psi_{R\pm }(0).
\end{equation}
Thus the two parts of the Kondo interaction are:
\begin{equation}
H_{int}=v_F\lambda_{++}\sum_{\pm}\psi^\dagger_{L\pm}{\vec \sigma \over 2}
\psi_{L\pm}\cdot \vec S
+ v_F\lambda_{+-}[\psi^\dagger_{L+}{\vec \sigma \over 2}
\psi_{L-}+\psi^\dagger_{L-}{\vec \sigma \over 2}
\psi_{L+}]\cdot \vec S .\label{asym}
\end{equation}
Initially
\begin{equation}
\lambda_{++}=\lambda_{+-}=\lambda = {2J\sin k_F\over \pi t}.
\end{equation}
However, while $\lambda_{++}$ remains marginal, even in the presence
of bulk interactions, $\lambda_{+-}$ does not.   The marginality of
$\lambda_{++}$ follows since the interaction can be written entirely
in terms of the spin bosons (on the two sides of the dot) whose
compactification radius (or interaction parameter, $g_\sigma$) remains
unchanged by the $SU(2)$ invariant bulk interactions.  On the other
hand, the dimension of $\lambda_{+-}$ changes with bulk interactions.
The corresponding operator has the same scaling dimension as a
standard hopping operator between two severed chains, Eq. (\ref{xsevered}).
Note that the spin-charge separation of the Kondo problem breaks down
once bulk interactions are included.  This results from the fact that
we can't transform to even and odd channels when bulk interactions are
present.  $\lambda_{+-}$ becomes irrelevant for repulsive interactions.
We expect a flow from the $\lambda =0$ fixed point either to the
resonant fixed point if the backscattering term is tuned to 0, or
otherwise to the charge and spin reflecting fixed point.  The length
scale at which the crossover to these other fixed points occurs is
again given by $\xi_K\propto e^{-1/\lambda_{++}}$ as before.  This
length scale should be the appropriate one to render the scaling
variable $Vl^{1/2-g_\rho /4}$, which controls the crossover between
resonant and charge/spin reflecting fixed points, dimensionless.
Thus we may write a general scaling form for the persistent current
in the presence of bulk interactions:
\begin{equation}
jl = f(\xi_K/l, (V/D)(l/\xi_K)^{1/2-g_\rho /4},\alpha ).\end{equation}
  Note in particular, that the 
width as a function of $V$ is given by $D(\xi_K/l)^{1/2-g_\rho /4}$.

\subsection{Asymmetric tunneling amplitudes}

In this section, we would like to study how the persistent currents are modified
if the tunneling amplitude between the wire and the dot
are non-symmetric. The tight-binding Hamiltonian (\ref{Handflux}) is replaced by:
\begin{equation} 
H=-t\sum_{j=1}^{l-2}(c^\dagger_jc_{j+1}+h.c.) 
-[c^\dagger_d(t_L e^{-i\alpha /2}c_{l-1}+t_R e^{i\alpha /2}c_1)+h.c.] 
   +\epsilon_dc^\dagger_dc_d 
   +Un_{d\uparrow}n_{d\downarrow},\label{Handflux1}
\end{equation} 
where we have introduced the left and right tunneling amplitudes $t_L,t_R$.
We have assumed no electronic interactions inside the ring. We refer to
section \ref{interact} for a discussion of this case, the idea being that one  more relevant operator  needs to be added to take into account the channel asymmetry in (\ref{asym}).
   
In the Kondo limit, the Hamiltonian (\ref{Handflux1}) reads: 
\bea 
H=-t\sum_{j=1}^{l-2}(c^\dagger_jc_{j+1}+h.c.) +2 \tilde{J}&&
\left[{t_L\over \sqrt{t_L^2+t_R^2}}~ e^{i\alpha /2}c^\dagger_{l-1}
+{t_R\over \sqrt{t_L^2+t_R^2}}~ e^{-i\alpha /2}c^\dagger_1 \right]
   {\vec \sigma \over 2}\cdot\vec S \nn\\ &&
\left[{t_L\over \sqrt{t_L^2+t_R^2}}~ e^{-i\alpha /2}c_{l-1}
+{t_R\over \sqrt{t_L^2+t_R^2}}~ e^{i\alpha /2}c_1 \right]
\label{Hkondoflux1}
\eea 
where the Kondo coupling constant has been defined by
$\tilde{J}=(t_L^2+t_R^2)({1\over-\eps_d}+{1\over U+\eps_d})$.
From (\ref{Hkondoflux1}), it is straightforward to reproduce our 
previous perturbative
calculations. The persistent current (\ref{jperte}) and (\ref{jperto}) simply become:
\bea
\tilde j_e(\alpha ) &=& {3\pi v_F e\over 4  l}\left[\kappa \sin \tilde \alpha 
[\lambda + \lambda^2\ln (lc)] +(1/4+\ln 2)\kappa^2\lambda^2\sin 
2\tilde \alpha \right] + O(\lambda^3,\kappa^3)\ \  (N \ \hbox{even}),\label{jperte1}\\
\tilde j_o(\alpha )&=&{3\pi v_F e\over 16  l}\kappa^2\sin 2\alpha [\lambda^2+2\lambda^3\ln 
(lc')]+O(\lambda^4,\kappa^4)\ \  (N\ \hbox{odd}),\label{jperto1} 
\eea
where we have introduced the ratio:
\beq
\kappa={2t_Rt_L\over t_L^2+t_R^2}.
\eeq
Notice first that the even and odd persistent currents are affected in a
different manner. More importantly, the infrared logarithm divergences are only
renormalizing the Kondo coupling as it should be. 
The persistent current is now a function of $\tilde \al,\lam_{eff},\kappa$.
The universal scaling form for $j$ (Eq. (\ref{univers})) is modified to:
\beq
j_{e/o}(\tilde \al,\kappa,\lam,l)={ev_F\over l} g_{e/o}(\xi_K/l,\tilde \al,\kappa)
\eeq
where the $g_i$'s are universal scaling functions. 
Our renormalization group
arguments prove that the  results obtained using perturbation theory 
are valid only when $\xi_K\gg l$. 

In order to analyse what happens in the limit
$\xi_K\ll l$, let us first observe  that the asymmetric (infinite length) lattice Hamiltonian:
\bea
H_{asym}=&&-t\sum\limits_{j\le -2} (c_j^\dag c_{j+1}+h.c.)-t\sum\limits_{j\ge 1}(c^\dag_j c_{j+1}+h.c.)\nn\\+
&&2 \tilde{J}
\left[{t_L\over \sqrt{t_L^2+t_R^2}}~ c^\dagger_{-1}
+{t_R\over \sqrt{t_L^2+t_R^2}}~ c^\dagger_1 \right]
   {\vec \sigma \over 2}\cdot\vec S 
\left[{t_L\over \sqrt{t_L^2+t_R^2}}~c_{-1}
+{t_R\over \sqrt{t_L^2+t_R^2}}~ c_1 \right]
\eea
has exactly the same continuuum limit 
(given by Eqs (\ref{cont},\ref{continuum},\ref{kinter})) 
as the symmetric Kondo Hamiltonian defined 
in Eq. (\ref{Hkondo}) (with $J\to \tilde J$), provided  we generalize the even-odd basis  in Eq. (\ref{psie}) to:
\bea
\psi_{e}(x)&=&{t_R\over \sqrt{t_L^2+t_R^2}}\psi_L(x)+{t_L\over \sqrt{t_L^2+t_R^2}}\psi_R(-x),\\
\psi_{o}(x)&=&{t_L\over \sqrt{t_R^2+t_L^2}}\psi_L(x)-{t_R\over \sqrt{t_L^2+t_R^2}}\psi_R(-x).
\eea
The generalized odd sector decouples from the generalized even sector which contains the Kondo interaction.
We can therefore follow the same reasoning as in section 2. The effective Kondo
coupling $\tilde\lam_1={4\tilde J \sin^2 k_F\over \pi v_F}$ renormalizes as in Eq. (\ref{renorm}). This effective dimensionless Kondo coupling is driven to strong coupling namely $\tilde\lam_1\to \infty$.
On the lattice, it means that a screening electron  forms a singlet with 
the impurity in the  orbital
defined by: 
\beq
c_e=({t_R\over \sqrt{t_R^2+t_L^2}} c_1+{t_L\over \sqrt{t_R^2+t_L^2}} c_{-1}).
\eeq 

In order to obtain the low energy effective Hamiltonian, we project
on the orthogonal  state $c_o$:
\bea
Pc_1P&=&{t_L\over \sqrt{t_R^2+t_L^2}}~c_o\nn\\
Pc_{-1}P&=&{-t_R\over \sqrt{t_R^2+t_L^2}}~c_o.
\eea
The low effective Hamiltonian then becomes:
\beq
H_{low}=-t\sum_{j\leq -3}(c^\dagger_jc_{j+1}+h.c.)-t\sum_{j\geq 
2}(c^\dagger_jc_{j+1}+h.c.)-t\left({t_L\over \sqrt{t_L^2+t_R^2}}
~c^\dagger_{2}c_o
-{t_R\over \sqrt{t_L^2+t_R^2}}~c_o^\dagger c_{-2}+h.c.\right)
.\label{Hlow1}
\end{equation} 
It is straightforward to show that this Hamiltonian has
the transmission probability:
\beq
T(k)=\left({2t_Rt_L\over t_R^2+t_R^2}\right)^2\sin^2k.
\eeq
According to section \ref{phb}, if $t_L,t_R$ are small enough meaning the bare Kondo coupling small enough, particle-hole
symmetry breaking effects are negligible. We can thus expect the transmission
probability at the Fermi surface to be given by $T(\eps_F)=\left({2t_Rt_L\over t_L^2+t_R^2}\right)^2$. A similar result
has been obtained in [\onlinecite{Glazman,Ng}]. In order to calculate the persisitent current, we can  directly apply the result of Gogolin and Prokof'ev.\cite{Gogolin} The persistent current
in the limit $\xi_K\ll l$ is therefore given by Eq. (\ref{gog}) with 
$T_F=T(\eps_F)$.

\section{Persistent currents for the side coupled quantum dot}
\label{pcsbqd}
We now want to consider the persistent current of a closed ring with
Aharonov-Bohm flux and a side coupled quantum dot.
As in the previous section, we suppose the ring has $l$ sites.
The Hamiltonian reads:
\beq
H=-t\sum\limits_{i=0}^{l-1} (c_{i+1}^\dag c_{i} +h.c.) + J~c_0^\dag { \vec{\s}\over 2}\cdot 
\vec{S}c_0. 
\eeq
There is no explicit flux dependence in the above Hamiltonian since we can choose  to gauge it away.
Obviously, the flux dependence is encoded into the non trivial 
boundary condition 
$c_{j+l}=e^{i\al}c_j$ and in particular $c_l=e^{i\al}c_0$, with $\Phi=(c/e)\al$. Due to this $\al$-dependent boundary condition, the possible values of the 
momentum are 
$$k_n=2\pi n/l+\al/l, ~n=-(l-2)/2,...,l/2~~({\rm l~even}).$$

When $J=0$,  the persistent current is the one of free fermions
given in Eq. (\ref{pcfree}) {\it except that $N$ has to be changed to $N-1$}.
Note that the persistent current is an odd function of $\al$ and is $2\pi$
 periodic for $N$ odd and $\pi$ periodic for $N$ even.
We will prove in the following that these relations persist to all order of the perturbation
scheme in the Kondo coupling $J$. 
Let us first distinguish the two cases $N$ even and $N$ odd.

\subsection{N even}

We first assume that $\al \in [0,\pi]$. 
The ground state is well defined and  has all levels empty or
doubly occupied except the Fermi level at $k_F$ defined by
\bea
k_F&=&{2\pi N\over 4l}-{\al\over l}, N/2~even\nn\\
k_F&=&{2\pi N\over 4l}-{(\pi-\al)\over l}, N/2~odd \label{fmom}
\eea
which contains one electron
forming a singlet with the impurity. 
Note that $k_F=k_F(\al)$.
The interaction is the standard Kondo interaction between the quantum dot and
the site $0$ of the ring.
There is a contribution at first order which does not depend on $\al$:
\beq
E_0^{(1)}=-{3J\over 4 l}.
\label{firstorder}
\eeq
The second order contributions can be obtained
according to  Eq. (\ref{e02}).
Following the procedure developed in appendix A, we first expand
$\psi(0)$ in Fourier modes and separate in the time ordered Green function 
the $k_F$ term (which
we call  $\psi_0$) from the rest of the series (which we call  $\psi'$).
 Due to $\alpha$ dependence of $k_F$, the left and
right branches gives different contributions to the propagator. 

Let us first suppose that $N/2$ is even. We will show how the case
$N/2$ odd is related to the case $N/2$ even later. 
We have depicted a schematic level diagram in figure \ref{disperse} for 
$\al \in [0,\pi]$ and $N=12$. The lowest energy level has momentum $k=\al/l$ corresponding to $n=0$. For $N/2$ even, the Fermi level is reached
by one electron with spin up or down belonging to the left branch (corresponding to $n=-3$ on the figure). 
\begin{figure}[h]
\hspace{3cm}\psfig{figure=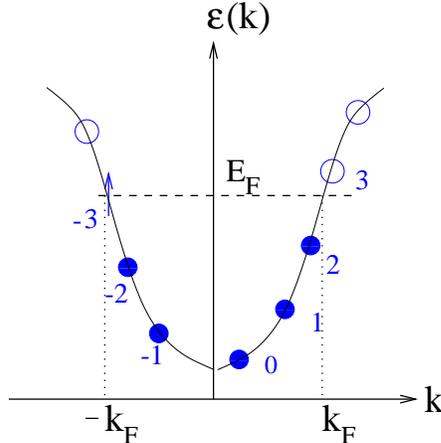,height=6cm,angle=0}
\caption{Level diagram for $N=12$ and $\al \in [0,\pi]$. The filled 
circles depict filled levels with two electrons whereas the empty circles are for empty levels.   The electron at the Fermi  level
 (n=-3) has a spin $\ua$ or $\da$.}
\label{disperse}
\end{figure}

From this diagram, it is straightforward to compute the  propagator
for the field $\psi'$ which are taken at site $0$: 
\beq
{\cal T}\la\psi_\al^{'\dag}(\tau)\psi_\be^{'}(0)\ra= \de_{\al,\be}  
G(\tau),
\eeq
with
\bea
G(\tau)&=& {1\over l}\sum\limits_{|k|<k_F} e^{(E_k-E_F)\tau} ~~{\rm if}~ \tau>0,\nn\\
&=&{1\over l}
\sum\limits_{m=1}^{l/2} (e^{\xi_m^L \tau}+e^{\xi_m^R \tau}),
\eea
where 
\bea
\xi_m^L&=&-2t\cos[2\pi (N/4-m)/l-\al/l]+2t\cos[2\pi N/4l-\al/l], \\
\xi_m^R&=&-2t\cos[2\pi (N/4-m)/l+\al/l]+2t\cos[2\pi N/4l-\al/l].
\eea
We   assume that at large $|\tau|$, the sum is dominated by excitations near the Fermi surface. We can therefore use the approximation $\xi_m^L\app 
-2\pi m v_F/l$ and $\xi_m^R\app 
-2\pi v_F(m-\al/\pi)$, such that the Green functions can be written as:
\beq
G(\tau)\app {1\over l}\left({e^{- {2\pi v_F\over l}\tau }\over 1-e^{- {2\pi v_F\over l}\tau} }
+{e^{- {2\pi\over l} v_F\tau (1-\al/\pi)} \over 1-e^{- {2\pi v_F\over l}\tau} }\right)\equiv G_1(\tau)~{\rm if}~ \tau>0.
\label{pro1}
\eeq
Similar calculations for $\tau<0$ leads to:
\bea
G(\tau)&=& -{1\over l}\sum\limits_{|k|>k_F} e^{(E_k-E_F)\tau} ~~{\rm if}~ \tau>0\nn\\
&\app& - {1\over l}\left({e^{ {2\pi v_F\over l}\tau }\over 1-e^{{2\pi v_F\over l}\tau} }
+{e^{  2 v_F\tau \al/l} \over 1-e^{ {2\pi v_F\over l}\tau} }\right)
\equiv -G_2(\tau)~{\rm if}~ \tau<0~.
\label{pro2}
\eea
The first part of the propagator comes from the left excitations whereas the
second part comes from the right excitations (see figure \ref{disperse}).

We now want to emphasize that  the only $\al$ dependence of the ground state energy comes from this propagator.
Thus, properties of this propagator extend to properties of the ground state and therefore of the persistent current.
First, notice that from time reversal symmetry it results that
 the ground state energy is an even function of $\al$ and
therefore the persistent current an odd function of $\al$.
 It is straightforward to show that
the case $N/2$ odd is obtained from the case $N/2$ even by changing $\al \to \pi-\al$ in (\ref{pro1}) and (\ref{pro2}). Note that it also a property of the
Fermi momenta defined in (\ref{fmom}). Using this property and 
time reversal invariance, 
we can prove
 that $E^{(0)}_{4p+2}(\al)=E^{(0)}_{4p}(\pi+\al)$. We can go one step further. 
Indeed, it is worth  noticing that under $\al \to \al+\pi$ the propagators
 (\ref{pro1}) and (\ref{pro2}) exchange with each other: $G_1(\tau)\to G_2(-\tau)$, $G_2(-\tau)\to G_1(\tau)$. $G_1$ is associated with particle excitations and $G_2$ wth hole excitations. When calculating the ground state energy in perturbation theory, each order of the perturbation theory appears as a time integral of a  function which should be symmetric under interchanging $G_1(\tau)$ with  $G_2(-\tau)$ due to particle-hole symmetry. This
proves  the $\pi$ periodicity of $E^{(0)}_{4p}(\al)$ and also
that: \beq E^{(0)}_{4p+2}(\al)=E^{(0)}_{4p}(\al).\eeq

We also need the correlator 
 for $\psi_0$:
\bea
{\cal T}\la\s|\psi_{0,\ga}^\dag(\tau)\psi_{0,\nu}(0)|\eps\ra &=&
{1\over l}[\de_{\s\ga}\de_{\eps \nu}- \th(-\tau)\de_{\ga\nu}\de_{\eps \s}],
\nn\\
{\cal T}\la\s|\psi_{0,\nu}(\tau)\psi_{0,\ga}^\dag(0)|\eps\ra &=&
{1\over l}[-\de_{\s\ga}\de_{\eps \nu}+\th(\tau)\de_{\ga\nu}\de_{\eps \s}],
\eea
where $|\eps\ra$ indicates an electron with spin $\eps$ lying at the Fermi level.
We have performed the calculations in the basis defined by: 
$$|A\ra,|B\ra=|\ua,\Da\ra,|\da,\Ua\ra,$$ 
for convenience. The double arrows denote the spin state of the impurity.
There are three contributions to the ground state energy at second order.
The first contribution comes from the
correlators involving only  $\psi'$ and reads:
\beq
-{J^2 l\over 4\pi v_F }\int\limits_0^\infty d u~ G_1(u)G_2(-u) 
Tr\left({\s^a\over 2}{\s^b\over 2}\right)[\la A|S^a(u) S^b(0)|B\ra+\la A| S^b(0) S^a(-u)|B\ra],
\eeq
where we have made the change of variable $\tau=2\pi v_F u/l$.
We need to add also the cross terms (containing two $\psi'$ and 
two $\psi_0$). 
The full result  for 
\beq
E_{AB}^{(2)}(\al)=-\dmi\int d\tau \la A|H_{int}(\tau)H_{int}(0)|B\ra,
\eeq
 reads:
\bea
E_{AB}^{(2)}(\al)&=& {-J^2l\over 4\pi v_F } \left[ 
\int\limits_0^\infty d u~( G_1(u)G_2(-u)+{1\over l}G_1(u))~Tr\left({\s^a\over 2}{\s^b\over 2}\right)~
[\la A| S^a(u) S^b(0)|B\ra+\la A| S^b(0) S^a(-u)|B\ra]\right.\nn\\
&-&{1\over l}\int\limits_0^\infty d u~
\left(G_1(u)\la A|c^\dag_{k_F}\left({\s^b\over 2}{\s^a\over 2}\right)
c_{k_F} S^a(u) S^b(0)|B\ra-G_2(-u)\la A|c^\dag_{k_F}\left({\s^b\over 2}{\s^a\over 2}\right)
c_{k_F} S^b(0) S^a(-u)|B\ra\right)\\
&+&{1\over l}\left. \int\limits_0^\infty d u~
\left(G_2(-u) \la A|c^\dag_{k_F}\left({\s^a\over 2}{\s^b\over 2}\right) 
c_{k_F} S^a(u) S^b(0)|B\ra-G_1(u)\la A|c^\dag_{k_F}\left({\s^a\over 2}{\s^b\over 2}\right) c_{k_F} S^b(0) S^a(-u)|B\ra\right)\right]\nn,
\eea
where we have also used $\tau=2\pi v_F u/l$.
Using the identity: 
$${\cal T}\la A|S^a(u)S^b(0)|B\ra={1\over 4}\de^{ab}\de_{AB}+{i\over 2}\eps(u)\varepsilon^{abc}\la A |S^c|B\ra,$$ and
gathering all contributions, we finally obtain the second order correction to the ground state energy:
\beq
E_0^{(2)}(\al)={-J^2l\over 4\pi v_F }  \int\limits_0^\infty du  \left[
{3\over 4}G_1(u)G_2(-u)+{9\over 8l}(G_1(u)+G_2(-u))\right].
\label{e2alpha}
\eeq
Note that this expression is symmetric under $G_1(u)\leftrightarrow G_2(-u)$ as it should be.
It is also worth noticing that had we just focussed on 
the leading logarithm divergences
in (\ref{e2alpha}), we would find $E^{(2)}(\al)\app -{3J^2\ln(lc)\over 4\pi v_F l}
=-{3\pi v_F\lam^2\ln(lc)\over 4l}$ which would renormalize perfectly the 
first order contribution to the ground state energy given in 
(\ref{firstorder}) (it also provides a non trivial check of the calculations). Nevertheless, in order to calculate the persistent current, we are only interested
in  the $\alpha$ dependent terms in (\ref{e2alpha}):
\beq
E^{(2)}(\al)\app {-J^2\over 4\pi v_F l}  \int\limits_0^\infty du  \left[
 {3\over 4}~{ [e^{(-2u+u\al/\pi)}+e^{-u(1+\al/\pi)}]
\over (1-e^{-u})^2 }+{9\over 8}~ {[ e^{-u\al/\pi}+e^{-u(1-\al/\pi)}]\over 1-e^{-u} } \right].
\eeq
 This expression contains some UV divergences which are removed when we 
consider the persistent current defined by 
$(j^e)^{(2)}=-e{d\over d\al}E^{(2)}(\al)$:

\beq
(j^e)_1^{(2)}(\al)= {ev_F\lam^2\over 4 l}  \int\limits_0^\infty du  \left[
 {3u\over 4}~{ [e^{(-2u+u\al/\pi)}-e^{-u(1+\al/\pi)}]
\over (1-e^{-u})^2 }-{9u\over 8}~ {[ e^{-u\al/\pi}-e^{-u(1-\al/\pi)}]\over 1-e^{-u} } \right].
\label{pceven}
\eeq
The subscript $1$ is for latter convenience.
This expression is completely antisymmetric under $\al\to \pi-\al$ as it 
should be (since $E_0(\al)=E_0(\pi-\al)$). 
In order to check explicitly the renormalization of the persistent current
with the Kondo coupling, 
we would need to go the third order in perturbation. 
We have not performed the third order calculation but we expect the dimensionless Kondo coupling constant $\lam={J\over \pi v_F}$
to renormalize as $\lam_{eff}(l)=\lam+\lam^2\ln l+...$ as in the previous section (see also further). 
Note that the $\al$ dependence of $k_F$ gives corrections smaller by a factor
of  $a/l$.
This expression contains some severe infrared divergences 
 when $\al\to 0$ or $\al\to \pi$. 
For example when $\al\to 0$, the  level at $k_F$ and the next higher level (
corresponding to  $n=3$ in Fig. \ref{disperse}) 
 become
 degenerate, which is not taken into account in our initial choice of the ground state.  
For small value of $\alpha$, we can evaluate the integral approximately by
\beq
(j^e)_1^{(2)}(\al) \app  -{9 e v_F\lam_{eff}^2\over 32 l} {\pi^2\over \alpha^2}.
\eeq
Physically, this corresponds to truncating the summation over intermediate 
states occurring
in second order perturbation to the first exited state only  which approaches
  the ground state as $\al \to 0$ and therefore gives the most important contribution to the persistent current.
For perturbation theory to make sense we are restricted to 
$(j^e_1)^{(2)}(\al)\ll (j^e)^{(0)}(\al)$) namely
\beq
\al\gg \pi\lambda_{eff} {3\over  \sqrt{32} }
\eeq
When this condition is not satisfied, this approach fails.
Note that this result is expected. Indeed, the persistent current is 
discontinuous at $\al=0$ for $\lam=0$ which precludes
 a naive perturbative analysis near this singular point which corresponds
 simply to a level crossing.
To overcome this difficulty occurring for   $\al\ll\pi$ or $|\al-\pi|\ll \pi$, we need to do perturbation theory around the correct ground state
which is built in this case from the two levels close to the Fermi surface
which are mixed by the perturbation. 
In the sequel we consider the case $\al\ll \pi$, which extends trivially to
$|\al-\pi|\ll \pi$ by symmetry around $\al=\pi/2$.
These two levels have momenta 
$k_{1/2}=\mp{2\pi N\over 4l}+ \al/l$. They correspond on the figure \ref{disperse} to the levels labeled by $n=-3$ and $n=3$.
In fact, the strategy we follow is analogous to  a second order 
degenerate perturbation theory. When $\al\ll \pi$,
the ground state is built with 
all levels with $|k|<k_1$  full and one electron lying on one of the 
two almost degenerate levels
forming a singlet with the impurity (defining therefore two possible states).

  The first order correction in the Kondo
coupling $J$ mixes 
these two states. It is straightforward to show that at this order the 
contribution of these two levels to the ground state energy
 is found by diagonalizing the $2\times 2$ matrix
\beq
\la H_{12}\ra=\left( 
\begin{array}{c c }
\eps_1-{3J\over 4l} & -{3J\over 4l}\\
 -{3J\over 4l} & \eps_2-{3J\over 4l}
\end{array} \right)~,
\label{mat}
\eeq
 written in the basis:
\beq
{|\ua,o;\Da\ra -|\da,o;\Ua\ra\over \sqrt{2}}~~;~~{|o,\ua;\Da\ra-|o,\da;\Ua\ra\over \sqrt{2}},
\label{singlet}
\eeq
where $|\ua,o;\Da\ra=c^{\dag}_{1,\ua}|\Da\ra$, 
$|o,\ua;\Da\ra=c^{\dag}_{2,\ua}|\Da\ra$, etc. 
Here the double arrows denote the spin
state of the impurity.

The other  levels are  giving  second order contributions in $J$
in (\ref{mat}). We have explicitly shown in appendix \ref{sbqde}   that the 
 Kondo coupling constant
in  Eq. (\ref{mat}) gets perfectly renormalized at second order,
 namely $J/l$ has to be replaced by
$\pi v_F(\lam+\lam^2\ln[lc])/l=\pi v_F\lam_{eff}(l)/l$. 
This result is not surprising  since we expect the persistent current to be a universal function of $\lam_{eff}$ and $\al$ as in Eq. (\ref{scale}).

Therefore the final result for the ground state energy 
can be cast in the form:
\bea
E^{(0)}&=&{-3\pi v_F \lam_{eff}\over 4l}+E^{free}(\al)+{\eps_2-\eps_1\over 2}-\sqrt{ {(\eps_1-\eps_2)^2\over 4}+{9\pi^2 v_F^2 \lam_{eff}^2\over 16l^2}},\\
&=&{-3\pi v_F\lam_{eff}\over 4l}+E^{free}(\al)+2t\sin[{2\pi N\over 4l}]\sin[\al/l]-\sqrt{ 4t^2\sin^2[{2\pi N\over 4l}]\sin^2[\al/l]
+{9\pi^2v_F^2\lam_{eff}^2\over 16l^2}},\nn
\eea
where $E^{free}(\al)$ is the ground state energy of the free case 
 for $N$ even. This is easily deduced from Eq. (\ref{pcfree}) by replacing
$N$ by $(N-1)$: 
\beq
E^{free}(\al)=constant+ {v_F\over 2\pi l}\left(\al^2+(\al-\pi)^2\right).
\eeq
For $\lam_{eff}=0$,  the free case is recovered as it should be.
The singularity in $\al=0$  is smoothed by the square root. 
The persistent current becomes so far in the large $l$ limit:
\bea
j_2^e(\al)&=&{-ev_F\over l}\left( {2\al\over \pi}-1+\cos[\al/l]-\dmi
{\sin[ 2\al/l] \over \sqrt{\sin^2[\al/l]+{9\pi^2\over 16 l^2}\lambda_{eff}^2}}
\right),\\
&\approx& {-ev_F\over l}\left(  {2\al\over \pi}- { {\al\over \pi} 
\over \sqrt{ ({\al\over \pi})^2+ ({3\lambda_{eff}\over 4})^2}}\right).
\label{pceven1}
\eea
\begin{figure}[t]
\hspace{3cm} \psfig{figure=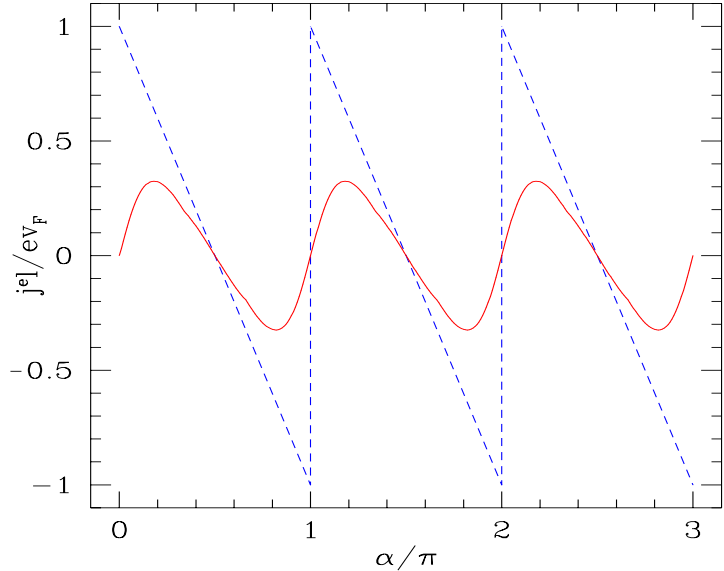,height=8cm,width=10cm,angle=0}
\caption{ Persistent current of the SCQD for $N$ even calculated  for $\xi_K/l\app 50$ (solid line) compared to the $J=0$ case (dashed line). }
\label{fpceven}
\end{figure}
Notice that for $\lam_{eff}\ne 0$, the persistent current is continuous in $0$
and $j_2^e(0)=0$.  
The same analysis can be reproduced 
near $\al=\pi$. Indeed, by symmetry it is enough to replace in (\ref{pceven1}) $\al/\pi\to 1-\al/\pi$ and to change the overall sign. 
It is clear that the result (\ref{pceven1}) is valid
only for small value of $\alpha\ll \pi$ and for $\lam_{eff}\ll 1$
 where the ground state can be regarded as
 a superposition of states belonging to
the two almost degenerate levels. For larger value of $\al\gg \lam_{eff}$ but still small
compared to $\pi$, we may evaluate the validity of this approach by developing (\ref{pceven1}) in powers of $\lam_{eff}/\al$:  
\beq
j_2^e(\al)\app(j^e)^{(0)}(\al) -{9 e v_F\lam_{eff}^2\over 32 l} {\pi^2\over \alpha^2}\app(j^e)^{(0)}(\al)+(j^e)_1^{(2)}(\al)~.
\eeq
In this regime, we recover the small $\al$ limit of (\ref{pceven}).
Thus Eqs (\ref{pceven}) and (\ref{pceven1}) agree in the limit 
$\lam_{eff}\ll \al\ll \pi$. Therefore, together, they cover all the $\al$ range at small $\lam_{eff}$.
We want to emphasize that these arguments are valid for small value of $\lam_{eff}\ll 1$ 
meaning $l\ll \xi_K$ where perturbation makes sense.

We have plotted the persistent current from Eq. (\ref{pceven}) and 
(\ref{pceven1})  (more exactly $j^e l/ev_F$)
in figure \ref{fpceven} for   $\xi_K/l \app 50$ 
($\lambda_{eff}\app 0.25$). We notice that the amplitude of the 
persistent current is already
strongly renormalized for this value of $\lam_{eff}$. Moreover the singularities at $\al =n\pi$ have been completely wiped out by the Kondo coupling. 

In the opposite limit $l\gg \xi_K$, where
perturbation breaks down, we may expect the persistent current to be given by the $J\to \infty$ limit according to standard renormalization group arguments
(see the discussion in section \ref{pceqd}). 
In this limit, we have shown in section 2 that the transmission $T$ tends toward $0$.
According to the argument given by Gogolin and Prokof'ev \cite{Gogolin}, we 
thus expect:   
\beq 
j^e\to 0 ~~{\rm for} ~~l\gg \xi_K,
\eeq
the picture being simply that the ring is ``cut'' into two pieces which prevents  non zero persistent currents.

\subsection{N odd}

The ground state is defined by all states
under the Fermi level  ($k_F=\pi(N-1)/4l$) being doubly occupied and all states above 
the Fermi level being empty.
The number of electrons inside the ring is even therefore 
the system, including the imputity, is in a doublet  state. Consequently, 
there is no contribution to the ground state at first order in $J$.
In the following, we   assume $N=4p+1$ and $0<\al<\pi$.
We will show later how the case $N=4p+3$ is related to that case $N=4p+1$.
We have depicted a schematic diagram of the different levels for $N=13$ 
in figure \ref{disperso}. The first level has a momentum $\al/l$ 
corresponding to $n=0$
and the last filled
level has momentum $k=(2\pi p -\al)/l$ (labeled by $-3$ in the figure \ref{disperso}). 

\begin{figure}[h]
\hspace{3cm}\psfig{figure=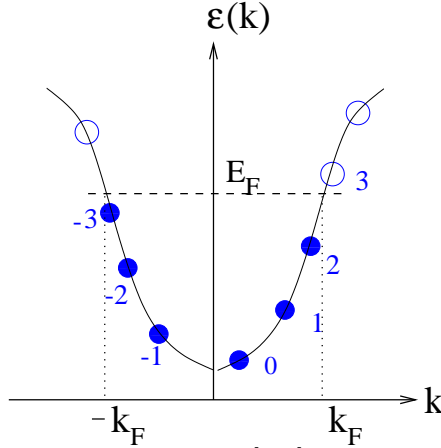,height=6cm,angle=0}
\caption{ Level diagram for $N=13$ and $\al\in [0,\pi]$. 
The filled  circles are for full levels whereas empty circles are for empty levels. The Fermi level $k_F$ lays
half between the highest filled level corresponding to $n=-3$ and the lowest empty level corresponding to $n=4$}
\label{disperso}
\end{figure}

The second order correction to the ground state can be written as
\beq
E_0^{(2)}=-\dmi \int\limits_{-\infty}^{+\infty}~ d\tau \ 
{\cal T} \la 0| H_{int}(\tau)H_{int}(0)|0\ra.
\eeq
The propagator reads:  
\beq
{\cal T}\la \psi_\al^\dag (\tau)\psi_{\beta}(0)\ra = \de_{\al,\be} G(\tau)=
{\de_{\al,\be}\over l}\sum_n
e^{\xi_n\tau}
\eeq
where 
\beq
\xi_n=-2t\cos[{2\pi\over l}n+{\al\over l}]+2t\cos k_F=
-2t\cos[{2\pi\over l}n+{\al\over l}]+2t\cos [{2\pi(N-1)\over 4l}]
\eeq
As for $N$ even , at large $\tau$, the sum is dominated by terms near 
the Fermi surface corresponding to $n={N-1\over 4} -m$, with $m\ll N$.
Due to the $\al$ dependence of the quantum numbers, there is an 
 asymmetry between 
left and right movers. By help of the level diagram \ref{disperso}, 
it can be easily shown that
\bea
G(\tau)&\app&{1\over l} \sum\limits_{m=1}^{l/2} e^{- {2\pi v_F\over l}\tau(m-\al/2\pi)}+{1\over l}
\sum\limits_{m=0}^{l/2} e^{- {2\pi v_F\over l}\tau(m+\al/2\pi)} \nn\\
&\app&{1\over l}\left( {e^{- {2\pi v_F\over l}\tau(1-\al/2\pi)}\over 1-e^{- {2\pi v_F\over l}\tau} }
+{e^{- ( v_F\tau \al/l)} \over 1-e^{- {2\pi v_F\over l}\tau} }\right)
\label{prop1}
\eea
This expression is valid whatever the sign of $\tau$.
Following similar arguments as for $N$ even, properties of the propagator
translate into properties of the ground state energy.
We can  show explicitly that $E_0(\al)=E_0(\al+2\pi)$
and especially $E_0(\al,4p+3)\equiv  E_0(\al+\pi,4p+1)$.
Therefore the case $N=4p+3$ is simply related to the case $N=4p+1$ by a translation of $\pi$.
The second order term in the ground state energy reads therefore:

\beq
E_0^{(2)}=-{J^2\over 2 }\sum\limits_{a,b} tr\left({\s^a\over 2}{\s^b\over 2}\right)
\la S^aS^b\ra \int\limits_{-\infty}^\infty G(\tau)[-G(-\tau)]
\eeq
After plugging the time ordered Green function (\ref{prop1}) in this expression 
we get, retaining only $\alpha$ dependent terms: 

\beq
E_0^{(2)}={3J^2\over 8 l (2\pi v_F)}\int\limits_0^\infty \ du ~{1+\cosh u(1-\al/\pi)
\over 1-\cosh u}.
\eeq

This result can be generalized to the third order in $J$.
Following similar calculations presented in appendix A2, we can show that:
\beq
E_0^{(3)}= -{3\over 16 } {J^3\over (2\pi v_F)^2} 4l\ln(lc)\int\limits_{-\infty}^{+\infty}
dv G(v)[-G(-v)].
\eeq

We define the Kondo dimensionless coupling constant $\lambda$ as $J=\pi v_F \lam$.  
Gathering all terms, we find
\beq
E_{0} \app  {-3\pi v_F\over 16 l}[\lam^2 +2\ln[lc]\lam^3]
\int\limits_{0}^{+\infty}
du ~{1+\cosh u(1-\al/\pi)
\over \cosh u-1}
\eeq
The correction to the persistent current finally takes the form :
\beq
(j^o)_1^{(2)}(\al) =  {-3 e v_F\lam_{eff}^2\over 16 l }\int\limits_0^\infty \ du ~{u\sinh u(1-\al/\pi)
\over \cosh u-1}~,
\label{pco1}
\eeq
where we have defined as usual $\lam_{eff}=\lam+\lam^2\ln(lc)$ the renormalized labeled Kondo coupling constant. Exactly as in the  case N even, this expression
 diverges  when $\al\to 0$, {\it i.e.} when the two levels close to the Fermi surface become  degenerate. In figure \ref{disperso}, it corresponds to the levels labeled by $n=-3$ and $n=3$.
  
For small value of $\alpha$, we can evaluate the integral approximately by
\beq
(j_1^o)^{(2)}(\al) \app  -{3 e v_F\lam_{eff}^2\over 16 l} {\pi^2\over \alpha^2}.
\eeq
Therefore, perturbation theory  applies when $(j^o)_1^{(2)}(\al)\ll (j_1^o)^{(0)}(\al)$ namely
\beq
\al\gg \pi\lambda_{eff} {\sqrt{3}\over 2 \sqrt{8} }.
\eeq
For smaller value of $\al \ll \pi$  
the Kondo coupling mixes strongly the states belonging to the 
two levels close to the Fermi surface. We need therefore 
to perform our perturbative
calculations around a new ground state built from these two almost 
degenerate levels which we denote  by  1 and 2. They  have   energy 
$\eps_{1/2}=-2t\cos(k_F\mp\al/l)$ respectively, where we have defined $k_F={2\pi (N-1)\over 4l}$. We can now repeat the same analysis as for $N$ even.

%

The ground state has all levels with $|k|<k_1$ full and
 two electrons in the levels $1,2$. We have  four possible states with 
$S_{tot}=S^z_{tot}=1/2$:
\beq
{|\ua,\da;\Ua\ra+|\da,\ua;\Ua\ra-2|\ua,\ua;\Da\ra\over \sqrt{6}}~~;~~|\ua\da,o;\Ua\ra~~;~~
|o,\ua\da;\Ua\ra~~;~~{|\ua,\da;\Ua\ra-|\da,\ua;\Ua\ra\over \sqrt{2}}~.
\label{basis}
\eeq
The states are defined according to $|\al,\be,\Ua\ra=c^\dag_{1,\al}c^\dag_{2,\be}|\Ua\ra$, $|\ua\da,o;\Ua\ra=c^\dag_{1,\ua}c^\dag_{1,\da}|\Ua\ra$, etc.
The first order contributions in $J$ mixes these states.
The associated matrix taking into account the contribution of these two levels
to the ground state energy at first order in $J$ reads:
\beq
\la H_{12} \ra =\left( 
\begin{array}{c c c c }
\eps_1+\eps_2-{J\over l} & -{J\sqrt{6}\over 4l} &  {J\sqrt{6}\over 4l}& 0\\
-{J\sqrt{6}\over 4l} & 2\eps_1 & 0 &0\\
 {J\sqrt{6}\over 4l} &0&2\eps_2&0  \\
0& 0& 0 & \eps_1+\eps_2 
\end{array} \right)~.
\label{matrix}
\eeq

We first note that the fourth state does not mix at this order and the matrix
is effectively $3\times 3$.
We want now to include the effects of the higher  levels
which  are  giving second order contributions in $J$
in (\ref{matrix}). We have shown in appendix \ref{sbqdo} using second order 
degenerate perturbation theory that the 
 Kondo coupling constant
in the Eq. (\ref{matrix}) gets perfectly renormalized namely $J/l$ has to be replaced by
$\pi v_F(\lam+\lam^2\ln[lc])/l=\pi v_F\lam_{eff}/l$.

Defining $E'=E-\eps_1-\eps_2-\sum\limits_{|k|<k_1}\eps_k$, 
the ground state energy is obtained by finding the minimum root of  the cubic  equation:
\beq
(E')^3+{\pi v_F\lam_{eff}\over l} (E')^2-E'\left[(\eps_1-\eps_2)^2+{3\pi^2v_F^2\lam_{eff}^2\over 4l^2}\right]-{\pi v_F\lam_{eff}\over l}(\eps_1-\eps_2)^2=0~.
\eeq
After the change of variable $E'={\pi v_F\over l}(X-{\lam\over 3})$, and using $\eps_2-\eps_1\approx
2v_F\al/l$, the equation can be cast
in the reduced form:
\beq
(X)^3-X[{13\over 12}\lam_{eff}^2+a^2]-{2\over 3}a^2\lam_{eff}+{35\over 108}\lam_{eff}^3=0~,
\eeq
where $a=2\al/\pi$.

The smallest root  is found to be
\beq
X_{min}=2\sqrt{Q}\cos[{\theta\over 3}+{2\pi\over 3}]~~~{\rm with}~~~~
\theta=\arccos\left[{-R\over Q^{{3\over 2}}}\right],
\eeq
where $Q,R$ are defined by:
\beq
Q={1\over 3}({13\over 12}\lam_{eff}^2+a^2)~~~,~~~R=-{1\over2}(-{2\over 3}a^2\lam_{eff}+{35\over 108}\lam_{eff}^3).
\eeq
Therefore, the ground state energy reads:
\beq
E^{(0)}=E^{free}+\eps_2-\eps_1+{\pi v_F\over l}X_{min}(2\al/\pi,\lam_{eff}),
\eeq
where $E^{free}$ is the ground state energy for $J=0$.

The persistent current can be then readily deduced as:
\beq
j_2^o\app-{2ev_F\over l} \left( {\hat{\al}\over \pi}+ X'_{min}(a=2\al/\pi,\lam_{eff})\right),
\label{pco2}
\eeq
with
$\hat\al=\al$ if $(N-1)/2$ is even and $\hat\al=\al-\pi$
 if $(N-1)/2$ is odd.
Moreover
\beq
X'_{min}={2a\over 3\sqrt{Q}}\cos[{\theta\over 3}+{2\pi\over 3}]-{2\sqrt{Q}\over 3}\sin[{\theta\over 3}+{2\pi\over 3}]\theta'(a)
\eeq
and 
\beq
\theta'(a)={a\lam_{eff}\over 3}~{
({29\over 24}\lam_{eff}^2-{1\over 3}a^2) \over 
Q^{5/2}\sqrt{1-{R^2\over Q^3}} }.
\eeq
Notice first that for $\lam=0$, the free case is recovered as it should be.
To analyze the range of validity of the formula (\ref{pco2}), it is interesting
to expand (\ref{pco2}) in $\lam_{eff}$. It is straightforward to show that:
\beq
j_2^o=(j^o)^{(0)}-{3 e v_F\lam_{eff}^2\over 16 l} {\pi^2\over \alpha^2}
\app (j^o)^{(0)}+(j_1^o)^{(2)},
\eeq
which corresponds the small $\al$ limit of the persistent current 
 calculated
with standard  perturbation theory (see Eq. (\ref{pco1})). 
Therefore Eqs (\ref{pco1}) and (\ref{pco2}) together cover all $\al$ for small
$\lam_{eff}$.
From both expressions (\ref{pco1}) and (\ref{pco2}), we have plotted the persistent current for $(N-1)/2$ even on figure \ref{pcodd} for $\xi_K/l\app 50$ ($\lam_{eff}\app 0.25$).

\begin{figure}[h]
\hspace{3cm} \psfig{figure=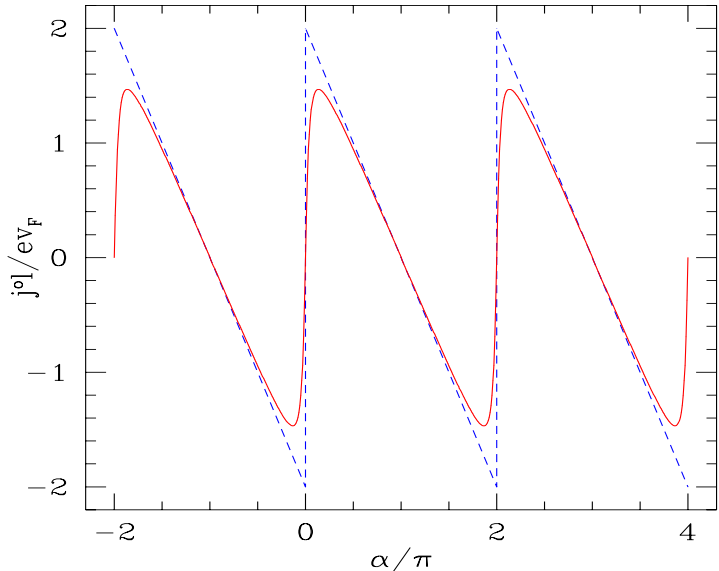,height=8cm,width=10cm,angle=0}
\caption{ Persistent current for $N=4p+1$ calculated for $\xi_K/l\app 50$ 
(solid line) versus the $J=0$ case (dashed line). The $N=4p+3$ case is obtained by a translation of $\pi$ of the horizontal axis. }
\label{pcodd}
\end{figure}

Notice that the corrections are weaker for $N$ odd that for $N$ even. 

In the opposite limit $\xi_K\ll l$, perturbation theory does not apply. 
However we
 can use the same reasoning as for $N$
even in order to prove that 
\beq 
j^o\to 0 ~~{\rm for}~~\xi_K\ll l, \eeq
 therefore no persistent current should be observed.

\section{Discussion and conclusion}

In the two previous sections, we have calculated the persistent current 
in the two limits $\xi_K/l\gg 1$ and $\xi_k \ll 1$ for both the EQD and SCQD.
Our results are summarized in figures \ref{empce},\ref{empco},\ref{fpceven},
\ref{pcodd}. 
In the embedded case, our results are actually very much different that
those obtained in [\onlinecite{Kang}] by solving some approximate 
self-consistent equations.  In particular, the current was predicted to be  small for $N$ odd {\it for large} $l/\xi_K$ but not  
for small $l/\xi_K$, the opposite of our result.
Furthermore the  $\pi$ periodicity in $\alpha$ that we find for odd $N$ was  
not obtained even in the limit of small $l/\xi_K$ where our perturbative
calculations are robust. This variational approach seems to give the inverse
behavior. We have obtained similar disagreement for the SCQD with ref.
[\onlinecite{Eckle}] and [\onlinecite{Cho}]. We refer to our previous paper \cite{letter} 
for a discussion of the validity of their results.

In this article, we have analyzed the validity and  robustness of our results against various perturbations which may occur experimentally.
Let us summarize our results and also discuss some other potential experimental
limitations like disorder.

We have first analyzed how the results are modified if we consider  electrons
 interactions inside the ring. From the theoretical point of view, particle-hole symmetry breaking terms are relevant for repulsive interactions. As a consequence, the  gate voltage controlling $\eps_d$ has to be tuned to resonance in order to reach a perfect transmittance.\cite{Kane} Therefore, it may imply 
experimentally that it might be difficult to reach the unitary limit and thus to observe the ``saw-tooth'' form of the  persistent current which is expected for $\xi_K\ll l$.
  
We have also treated the case of asymmetric tunneling amplitudes $t_L\ne t_R$
between the EQD  and the wires. This affects especially the strong coupling regime $l\gg \xi_k$ because the transmission probability is no longer $1$ and therefore
the persistent current  loses its ``saw-tooth'' shape. 
Such effect could disguise the cross-over between the regimes $l\ll \xi_K$
to $l\gg \xi_K$. From the experimental point of view, it would be important 
to tune the  voltage gates $V_l$ and $V_R$ (see figure \ref{dots}) associated 
with $t_L$ and $t_R$ as close as possible to each other in order to observe such cross-over. Moreover, in the presence of repulsive interactions, the operator
corresponding to the tunneling asymmetry  becomes relevant at resonance. It would therefore imply experimentally one more parameter to tune.

We have supposed the wire to be relatively clean such that
we can neglect the effect of non-magnetic impurities.
One may wonder how our results are modified when we have both the quantum dot and just one non-magnetic impurity with a potential scattering $V$ localized at a distance $r$ from the quantum dot. In the regime $r<l\ll \xi_K$, we can easily generalize our perturbative calculations for both the EQD and SCQD. We expect  the persistent current to be given
by a scaling function of $\al,\xi_k/l,r/l,V$ in this limit:
\beq
jl=g(\al,\xi_k/l,r/l,V),
\eeq
where $g$ is a universal function depending on the system we consider
(EQD versus SCQD) but also on the parity of the system. We have explicitly checked for the EQD that the first correction to the persistent current is 
$O(JV^2)$ for $N$ even and  $O(J^2V^2)$ for $N$ odd.
In the other limit, $l>r\gg \xi_K$, it is reasonable to expect that the persistent current can be obtained by simply taking the limit $J\to \infty$.
We can then easily calculate the transmission probability
of the associated non interacting model.
At half filling, the transmission probability reduces to the one of the non-magnetic impurity: $T=v_F^2/(v_F^2+V^2)$ (away from half filling, the expression is more complicated and has some $r$ dependence). Plugging this result into (\ref{gog}), provides
the expression for the persistent current.
The outcome is a decrease in the amplitude of the persistent current but especially the loss of the ``saw-tooth'' form here too. Such a result might also disguise a clear cross-over from $\xi_K\gg l$ to $\xi_K\ll l$ if $V$ is strong enough. From the experimental point of view, it seems quite important that the wires
should be relatively clean. The fact that the unitary limit has been reached in
[\onlinecite{Wiel}] may indicate  indeed that the wires are clean enough
such that these effects are negligible.

We have assumed only one electronic channels but despite the narrowness
of present quantum wires, several electronic channels can be activated. 
  However, it is reasonable to expect that one channel 
will have a stronger tunneling amplitude to the dot than the 
others.  The RG equations, to third order, for the multi-channel 
Kondo problem are: 
\begin{equation} 
d\lambda_i/d\ln l = \lambda_i^2-(1/2)\lambda_i\sum_j\lambda_j^2. 
\label{RG3}\end{equation} 
We see that if all but one of the couplings, $\lambda_1$, are small, while  
$\lambda_1$ is larger and positive, then we may approximate the equation  
for the small couplings by only the second term in Eq. (\ref{RG3}), keeping  
only the $\lambda_i\lambda_1^2$ term.  This equation then predicts  
that all the small couplings shrink.  Meanwhile, the larger 
coupling grows. 
Therefore, the impurity spin is screened by an electron from  
the most strongly coupled channel and the other channels decouple at  
low energies.
 Thus, for $l>>\xi_K$, we expect the single channel result 
obtained for the embedded quantum dot to still apply. 
Yet, we expect the results for the side coupled quantum dot to be fairly 
different. Indeed, if just one channel is screened by the spin impurity,
the other channels do not feel the effects of the quantum dot in the
scaling limit. The larger the number of channels,
the less the transmission is affected by the dot.
Therefore, if there are several active channel in a side coupled dot 
experiment, the observation of the cross-over between large $\xi_K/l$ to small $\xi_K/l$
could be severely concealed.
In this respect, the embedded quantum dot would be a better candidate
to detect the screening cloud through persistent current measurements.

In this paper we have shown that the persistent current in a mesoscopic
ring coupled to a quantum dot is a highly sensitive function of the ratio
of the ring circumference to the size of the Kondo screening cloud.  On
the other hand, electron interactions in the ring, asymmetric tunnelling
amplitudes, disorder and the presence of several channels can all serve to
mask this sensitivity. Thus, while the screening cloud size is an
important length scale in quantum dot physics, its separation from various
other effects is likely to be a challenging experimental problem.

\acknowledgements
We would like to acknowledge interesting discussions with N. Andrei and 
S. De Franceschi. 
This research was supported in part by NSERC of Canada.

\appendix 
\section{Perturbation theory for the embedded quantum dot} \label{eqd}
 \subsection{$N$ even} \label{eqde}
In this appendix, we calculate the ground state energy at second order in perturbation theory for N even. The correction to the ground state energy is 
given by:
\beq
E_0^{(2)}=-{J^2\over 2!}\int\limits_{-\infty}^{+\infty}~d\tau~{\cal T}\la s|
\chi^\dag(\tau){\vec \s \over 2}\cdot\vec S~\chi(\tau)
\chi^\dag(0){\vec \s \over 2}\cdot\vec S~\chi(0)|s\ra,
\eeq 
where $\chi$ has been defined in (\ref{defchi}).
The first contribution to $E_0^{(2)}$, coming from the 
term involving 2 factors of $\chi {'}$ in $H_{int}$, reads: 
\begin{equation} 
-{1\over 2!}J^2\hbox{tr}\left({\sigma^a\over 2}{\sigma^b\over 2}\right) 
<s|S^aS^b|s>\int_{-\infty}^\infty d\tau G(\tau )^2, 
\end{equation} 
where $G(\tau)$ has been defined in (\ref{G(tau)}).
To evaluate the contribution of the cross-terms between $\chi_0$ and 
$\chi {'}$ we need: 
\begin{eqnarray} 
{\cal T}<\ga |\chi_{0,\eps}^\dagger (\tau 
)\chi_{0,\nu}(0)|\beta > &=&{4\over l}\sin^2k_F(1-\cos \tilde 
\alpha )[\delta_{ \gamma\eps} \delta_{\beta \nu}-\theta (-\tau 
)\delta_{\eps \nu}\delta_{\beta 
\ga}]\nonumber \\ 
{\cal T}<\ga |\chi_{0,\nu} (\tau )\chi_{0,\eps}^\dagger 
(0)|\beta > &=&{4\over l}\sin^2k_F(1-\cos \tilde \alpha 
)[-\delta_{\gamma\eps} \delta_{\beta \nu}+\theta (\tau 
)\delta_{\eps \nu}\delta_{\beta \ga}], \label{propa2}
\end{eqnarray} 
where 
$|\ga\ra$ labels the state with one electron of spin $\ga$ at 
the Fermi level.  The full result for the ground state energy to 
 second order in $J$ can be written: 
\begin{eqnarray} 
E_0^{(2)}&=&-{J^2\over 2!}\int_{-\infty}^\infty d\tau\sum_{a,b} 
\Biggl\{ \hbox{tr}{\sigma^a\over 2}{\sigma^b\over 2}<s|S^aS^b|s> 
\left[G^2(\tau )+{4\over l}\sin^2k_F (1-\cos \tilde \alpha )G(\tau 
)\epsilon (\tau ) \right]\nonumber \\ && +{4\over 
l}\sin^2k_F(1-\cos \tilde \alpha )<s|c^\dagger_{k_F}\left[ 
{\sigma^a\over 2},{\sigma^b\over 2}\right] c_{k_F}{\cal T}[ 
S^a(\tau )S^b(0)|s> G(\tau )\Biggr\}.\label{E021}\end{eqnarray} Here we have 
reinserted the creation operator at the Fermi surface, $c_{k_F}$ 
(at $\tau =0$) in order to facilitate comparison with the first 
order term.  Now using the impurity spin Green's function: 
\begin{equation} 
{\cal T}<S^a(\tau )S^b(0)> = {1\over 4}\delta^{ab}+{i\over 2} 
\epsilon (\tau )\epsilon^{abc}S^c,\end{equation} 
this can be simplified to: 
\begin{eqnarray} 
E_0^{(2)}&=&-{J^2\over 2!}\int_{-\infty}^\infty d\tau \Biggl\{ 
{3\over 8}\left[G^2(\tau )+{4\over l}\sin^2k_F(1-\cos \tilde 
 \alpha )G(\tau )\epsilon (\tau ) 
\right]\nonumber \\ && 
 -{4\over l}\sin^2k_F(1-\cos \tilde \alpha )<s|c^\dagger_{k_F} 
{\vec \sigma \over 2}c_{k_F}\cdot \vec S|s>G(\tau )\epsilon (\tau )\Biggr\}. 
\label{E}\end{eqnarray} 
 
We now wish to examine the behavior of these various terms at large $l$. 
In the limit $l>>v_F|\tau |$, 
\begin{equation} 
G(\tau ) \approx {4\sin^2k_F\over \pi v_F\tau }.\end{equation} To 
estimate the large -$l$ behavior of $\int d\tau \epsilon (\tau ) 
G(\tau )$ we may use this expression for $\tau_0<|\tau |<l/v_F$ 
where $v_F\tau_0$ is of order  the lattice constant (1 in our 
units).  At larger $\tau$, $G(\tau )$ decays to zero 
exponentially, as we see from Eq. (\ref {G(tau)}). Thus, 
\begin{equation} 
\int_{-\infty}^\infty d\tau \epsilon (\tau )G(\tau ) \to 
{8\sin^2k_F\over \pi v_F}[\ln lc_1
+\cos\tilde \alpha \ln 2] 
,\end{equation} 
where $c_1$ is a constant of  $O(1)$, independent of $l$ at large $l$. Now consider: 
\begin{equation} 
\int_{-\infty}^\infty G(\tau )^2 \approx {16\sin^4k_F\over 
l^2}\int d\tau  \left[ {1\over 
 (e^{\pi v_F|\tau |/l}-1)^2}+ 
{2\cos \tilde \alpha \over (e^{2\pi v_F|\tau |/l}-1)} 
+{\cos^2\tilde \alpha \over  (e^{\pi v_F|\tau |/l}+1)^2}\right]. 
\end{equation} 
The integral of the first term is divergent at $\tau =0$.  This just 
reflects our use of the large $|\tau |$ form of $G(\tau )$.  Using the 
exact expression would give an integral of $O(l^2)$, corresponding to 
a term of $O(1)$ in the ground state energy.  This is independent of 
$\alpha$ however, and so  does not contribute to the persistent 
current, using Eq. (\ref{current}).  Thus we have: 
\begin{equation} 
\int_{-\infty}^\infty G(\tau )^2 \to {16\sin^4k_F\over l}\left[ 
2\cos \tilde \alpha \left({\ln l\over \pi v_F} 
+\hbox{finite}\right) + {\cos^2\tilde \alpha \over \pi v_F}(-1+2\ln 
2) \right] + \hbox{constant}.\end{equation} Importantly, the $\ln 
l$ terms in the term proportional to tr$\sigma^a\sigma^b$, in Eq. 
(\ref{E021}),  cancel.  Now evaluating the various terms in Eq. 
(\ref{E}), we have: 
\begin{eqnarray} 
E_0^{(2)} \to -{J^2\sin^4k_F\over 2!l\pi v_F}\Biggl\{ 
\left[ \cos \tilde \alpha \times \hbox{finite} -6\cos^2\tilde 
\alpha \right]\nonumber \\ 
 -(1-\cos \tilde \alpha ) 
32(\ln lc_1 +\ln 2\cos\tilde \alpha )<s|c^\dagger_{k_F} {\vec \sigma \over 
2}c_{k_F}\cdot \vec S|s>\Biggr\}+\hbox{constant}. 
\end{eqnarray} 
Combining this with the term of $O(J)$, gives: 
\begin{equation} 
E_0 = {3\pi v_F\over 4l}\left[ \cos \tilde \alpha 
[\lambda+\lambda^2 \ln (lc)]+\cos^2\tilde \alpha (1/4+\ln 2) 
\lambda^2\right]+\hbox{constant}, 
\end{equation} 
where $c$ is a dimensionless constant which we have not calculated 
explicitly, and $\lambda$ is the dimensionless Kondo coupling 
defined in Eq. (\ref{lambda1def}).
 
\subsection{ $N$ odd} \label{eqdo}
Let us
write in this case an expression for the term in the ground state energy of 
third order in $J$: 
\begin{eqnarray} 
\beta E_0^{(3)}& =& {J^3\over 3!}2\sum_{a,b,c}\hbox{tr}\left( {\sigma^a\over 2} 
{\sigma^b\over 2}{\sigma^c\over 2}\right){\cal T}<S^a(\tau_1) 
S^b(\tau_2)S^c(\tau_3)>\nonumber \\ 
&&\cdot \int d\tau_1d\tau_2d\tau_3 G(\tau_2-\tau_1) 
G(\tau_3-\tau_2)G(\tau_1-\tau_3).\end{eqnarray}  The factor of $2$ arises 
from the two ways of ordering the three vertices. 
Here $\beta$ is the inverse temperature which must be taken to $\infty$. 
The integrals run between $\pm \beta /2$.  We now use: 
\begin{equation} 
{\cal T}<S^a(\tau_1)S^b(\tau_2)S^c(\tau_3)>={i\over 8}\epsilon^{abc} 
\epsilon (\tau_1,\tau_2,\tau_3),\end{equation} 
 where $\epsilon (\tau_1,\tau_2,\tau_3)=1$ if $\tau_1>\tau_2>\tau_3$ 
and is completely antisymmetric in its arguments.  Using Eq. 
(\ref{G(tauodd)}) for $G(\tau )$, we now get four terms proportional 
to $\cos^n \alpha$ with $n=0$, $1$, $2$ or $3$.  The $n=0$ term does 
not contribute to the persistent current so we ignore it.  The $n=1$ 
and $n=3$ terms vanish because the integrands are antisymmetric. 
This leaves only the $\cos^2\alpha$ term, which is: 
\begin{equation} 
E_0^{(3)} \approx {-3J^3\cos^2\alpha 
\over 2l^3}\sin^6k_F\int_{-\infty}^\infty d\tau_1d\tau_2 
{\epsilon (\tau_1,\tau_2,0)\over \cosh (\pi \tau_1v_F/2l) 
\cosh (\pi \tau_2v_F/2l)\sinh [\pi (\tau_2-\tau_1)v_F/2l]}. 
\end{equation} 
This integral is actually divergent at $\tau_1=\tau_2$, signaling 
the breakdown of the large $\tau$ approximation to $G(\tau )$. 
Integrating over $v_F|\tau_1-\tau_2|\leq l$ gives a contribution 
to the energy proportional to $\ln l$.  We may set $\tau_2$ equal 
to $\tau_1$ inside the argument of the (non-singular) $\cosh$ 
function, so that the logarithmic term can be obtained as: 
\begin{equation} 
E^{(3)}_0 \approx {3J^3\cos^2\alpha \over 2l^3}\sin^6k_F\int 
d\tau_1{1\over \cosh^2(\pi v_F\tau_1/2l)} \int d\tau {\epsilon 
(\tau )2l\over v_F(\tau )\pi}, 
\end{equation} 
where the $\tau$ integral should be taken over $\tau_0<|\tau |<l/v_F$, 
where $\tau_0v_F$ is a distance of order a lattice spacing (one in 
our units). Thus we obtain: 
\begin{equation} 
E^{(3)}_0 = {24J^3\cos^2\alpha \sin^6k_F\ln (lc')\over (v_F\pi 
)^2l} + \hbox{constant},\end{equation} where $c'$ is another 
constant of O(1) which we have not determined. Combining this with 
the second order result, of Eq. (\ref{E02odd}) we obtain: 
\begin{equation} 
E_0=\cos^2\alpha {3\pi v_F\over 16l}[\lambda^2+2\lambda^3\ln (lc)] 
+ \hbox{constant}. 
\end{equation} 

\section{$J\to \infty$ transmission probability in the embedded quantum dot} 
We can easily calculate the exact transmission coefficient for the 
model with $\alpha +\pi =0$ in the Eq. (\ref{Hlowalpha}).
 We simply solve the lattice 
Schroedinger equation: 
\begin{eqnarray} 
E\phi_j&=&-t(\phi_{j+1}+\phi_{j-1}),\ \  (j=3,4,\ldots 
l-3)\nonumber \\ 
E\phi_2&=&-(t/\sqrt{2})\phi_a-t\phi_3 \nonumber \\ 
E\phi_{l-2}&=& -t\phi_{l-3}-(t/\sqrt{2})\phi_a \nonumber \\ 
E\phi_a 
&=&-(t/\sqrt{2})(\phi_2+\phi_{l-2}).\label{Sch}\end{eqnarray} The 
phase shift results from the reduction of the hopping matrix 
element by a factor of $1/\sqrt{2}$ at the site $a$, again a 
consequence of the projection onto the odd state.  We can write 
down the general solution of Eq. (\ref{Sch}),  by inspection: 
\begin{eqnarray} 
\phi_j &=& A\epsilon (j)\sin k(|j|-1)+B\sin k|j|,\ \ (|j|\geq 2),\nonumber \\ 
\phi_a&=& B\sqrt{2}\sin k,\label{wf}\end{eqnarray} 
with $\eps(j)=1$ if $j>0$ and $-1$ if $j<0$.
In Eq. 
(\ref{wf}), we label sites to the left of the impurity by negative 
integers, so $\phi_j$ is the wave-function at site $l+j$, with 
$j=-2,-3,\ldots$. The allowed values of $k$ depend on $l$ but we 
don't need to consider them explicitly since we just need the 
transmission probability. 
 For suitably chosen $A$ 
and $B$ we can obtain a scattering solution of the form: 
\begin{eqnarray} 
\phi_j&=&e^{ikj}+\sqrt{R}e^{i\delta_b}e^{-ikj},\ \  (j\leq -2)\nonumber \\ 
\phi_j &=& \sqrt{T}e^{i\delta_f}e^{ikj},\ \  (j\geq 
2),\end{eqnarray} with 
\begin{equation} 
T(k)=\sin^2k.\end{equation} 
 
\section{Perturbation theory for the side coupled quantum dot}

\subsection{Degenerate perturbation theory for N even}\label{sbqde}
When $\al\ll \pi$,
we need to perform our perturbative calculations around a different 
ground state which is composed of the two levels near $k_M=\pi N/4l$
(we have used $k_M$ which is different from $k_F$ defined in Eq. 
(\ref{fmom})). 
In this section, 
we want to include  higher levels in Eq. (\ref{mat}) at second order
in perturbation.
We make the approximation that the levels $1$ and $2$ are almost degenerate
with energy $\eps_1\approx \eps_2\app\eps_M$. 
We proceed strictly speaking with a second order degenerate perturbation theory.
In other words, when  introducing the sum over intermediate states (with energy $\eps_k$)
 in the standard  second order perturbation
formula,  we are approximating $\eps_k-\eps_{1/2}$ as $\eps_k-\eps_M$.
We expect this approximation to be valid as far as $|\eps_{1/2}-\eps_M|\ll |\eps_k-\eps_M|$ which implies $\al\ll\pi$. Obviously $\lam$ is supposed
to be small compared to $1$ for perturbation to make sense. 
There are no contribution at first order in $\lam$ from the other levels.
The first contribution comes at second order:
\beq
E^{(2)}_{pq}=-\dmi \int d\tau \la p| {\cal T}H_{int}(\tau)H_{int}(0)|q\ra
\eeq
where $p,q=1,2$ labels the two different singlet states (see Eq. (\ref{singlet})). Note that in our approximation, we suppose that the states $|p\ra$ and $|q\ra$ have the same 
 energy $\eps_M$ which is also {\it a priori} required for this formula to make sense.
In the mode expansion of $\psi$, we separate as before $\psi_{1/2}$ (corresponding 
to momenta $k_{1/2}$) from other modes (which we call  $\psi'$).
The propagator for the $\psi'$ field  remains unchanged:
\beq
{\cal T} \la\psi^{'\dag}(\tau)\psi'(0)\ra={\veps(\tau)\over l}\ {2\cosh(v_F\tau\al/l)\over e^{2\pi v_F|\tau|/l}-1}=G(\tau)~.
\eeq

We also need the other propagators:
\bea
&&{\cal T} \la i{\ga}|\psi^{\dag}_{k,\eps}(\tau)\psi_{k',\de}(0)|j{\be}\ra
={1\over l}[\de_{kk_i}\de_{k'k_j}\de_{\eps\ga}\de_{\de\be}
-\Theta(-\tau)\de_{kk'}\de_{k_ik_j}\de_{\eps\de}\de_{\ga\be}]\\
&&{\cal T} \la i\ga|\psi_{k',\de}(\tau)\psi_{k,\eps}^\dag(0)|j\be\ra
={1\over l}[-\de_{kk_i}\de_{k'k_j}\de_{\eps\ga}\de_{\de\be}
+\Theta(\tau)\de_{kk'}\de_{k_ik_j}\de_{\eps\de}\de_{\ga\be}]~,
\eea
whith $k,k'=k_1,k_2$, and where the electronic state $|i\ga\ra$ corresponds to on electron with spin $\ga$ with
momentum $k_i=k_{1/2}$.

It can then be shown easily that
\bea
E^{(2)}_{pq}&=&-{J^2\over 2 }\int\ d\tau\sum_{a,b}\left\{ tr{\s^a\over 2}{\s^b\over 2}<p|S^aS^b|q>
\left[ G^2(\tau)+{2\over l} G(\tau)\veps(\tau)\right]\right. \nn\\
&+&{1\over l}G(\tau)~\left. \la p|c^\dag_{k_p} \left[{\s^a\over 2},{\s^b\over 2}\right]c_{k_q}
{\cal T}\la p|S^a(\tau)S^b(0)|q\ra .\label{epq}
\right\}
\eea
Notice that the $\log(l)$ terms in the  first line of Eq. (\ref{epq}) cancel.
We  find that the leading behavior at large $l$ is: 
\beq
E^{(2)}_{pq}=- {3J^2\over 4 \pi v_F l}\ln lc=- {3\pi v_F\lam^2\over 4  l}\ln lc,
\eeq
which is independent of $p,q$ and
 renormalizes perfectly the matrix elements of Eq. (\ref{mat}).
We have therefore proven that in the range   $\alpha\ll \pi$, the main 
effect of the  higher  levels is to renormalize the Kondo coupling, the infrared divergences being cut off by the size of the ring.

\subsection{Degenerate perturbation theory for N odd}\label{sbqdo}

When $\al\ll \pi$,
the two levels near $k_F=\pi(N-1)/4l$ become almost degenerate (see Fig. \ref{disperso}).
We want now to include the effects of the other levels in the Eq. 
(\ref{matrix}). 
They are contributing to the kinetic energy but are also giving second order contributions in $J$
in (\ref{matrix}). We follow exactly the same scheme  as for N even. Namely, we assume that the two levels $1$ and $2$ close to the Fermi surface are almost degenerate and have energy $\eps_{1/2} =-2t\cos[k_F]$.
By analogy with second order degenerate perturbation theory,
we need to compute:
\beq
E^{(2)}_{ij}=-\dmi \int d\tau \la i|{\cal T} H_{int}(\tau)H_{int}(0)|j\ra.
\label{degpert}
\eeq
where $1\le i,j\le 4$ correspond to the four states defined in (\ref{basis})
which are taken to have the same ground state energy $\eps_i\app\eps_F$ for (\ref{degpert}) to make sense. 
It seems more convenient to first compute the quantities $A^{(2)}_{kl}=-\dmi \int d\tau \la S_k| H_{int}(\tau)H_{int}(0)|S_l\ra $ where the states $|S_k\ra$ define the basis
$$|S_1\ra=|\ua,\ua;\Da\ra~;~|S_2\ra=|\ua,\da;\Ua\ra~;~|S_3\ra=|\da,\ua;\Ua\ra~
;~|S_4\ra=|\ua\da,o;\Ua\ra~;~|S_5=\ra|o,\ua\da;\Ua\ra$$

As usual, we separate in the Fourier decomposition of $\psi_j={1\over \sqrt{l}}
\sum_k e^{i k j} \psi_k$ the modes involving $k_{1/2}=k_F\mp\al/l$ from other modes, which we label by
 $\psi'$.
The propagator involving the prime fields only is easily performed:
\beq
{\cal T} \la\psi^{'\dag}(\tau)\psi'(0)\ra={\veps(\tau)\over l}\ {2\cosh(v_F\tau\al/l)\over e^{2\pi v_F|\tau|/l}-1}=G(\tau)~.
\label{prop}
\eeq
Let us define the electronic   states $|r_i\ra$ and $\la r_j|$ as follow:
 $$|r_i\ra=\psi_{p_1,\ga_1}^\dag\psi_{p_2,\ga_2}^\dag|0\ra~~~;~~~
\la r_j|=\la0|\psi_{q_2,\be_2}\psi_{q_1,\be_1}~,$$
where the momenta $p_1,p_2,q_1,q_2$ take the values $k_1$ or $k_2$ and $\ga_1,\ga_2,\be_1,\be_2$ are 
the associated spins (the vacuum $|0\ra$ means the Fermi sea i.e which has levels with $|k|< k_1$ full). Let us also define the impurity state $|u_i\ra$ such
that $$|S_i\ra=|r_i\ra \bigotimes |u_i\ra$$ 

The other propagators we need are the following:
\bea
&&{\cal T} \la r_i|\psi^{\dag}_{k,\eps}(\tau)\psi_{k',\nu}(0)| r_j\ra ={\cal T}
\la 0|\psi_{q_2,\be_2}\psi_{q_1,\be_1}\psi^{\dag}_{k,\eps}(\tau)\psi_{k',\nu}(0)\psi_{p_1,\ga_1}^\dag\psi_{p_2,\ga_2}^\dag|0\ra\nn\\
&=&{1\over l}\left[ \de_{k'p_1}\de_{\nu\ga_1}(\de_{kq_1}\de_{p_2q_2}\de_{\eps\be_1}\de_{\ga_2\be_2}
-\de_{kq_2}\de_{p_2q_1}\de_{\eps\be_2}\de_{\ga_2\be_1})\right.\nn\\&&
+\de_{k'p_2}\de_{\nu\ga_2} (\de_{kq_2}\de_{p_1q_1}\de_{\eps\be_2}\de_{\ga_1\be_1}-
\de_{kq_1}\de_{p_1q_2}\de_{\eps\be_1}\de_{\ga_1\be_2})
-\left.\Theta(-\tau)\de_{kk'}\de_{\eps\nu}\de_{s_is_j}\right] 
\eea
\bea
&&{\cal T} \la r_i|\psi_{k',\nu}(\tau)\psi^{\dag}_{k,\eps}(0)| r_j\ra ={\cal T}
\la 0|\psi_{q_2,\be_2}\psi_{q_1,\be_1}\psi_{k',\nu}(\tau)\psi^\dag_{k,\eps}(0)\psi_{p_1,\ga_1}^\dag\psi_{p_2,\ga_2}^\dag|0\ra\nn\\
&=&{1\over l}\left[- \de_{k'p_1}\de_{\nu\ga_1}(\de_{kq_1}\de_{p_2q_2}\de_{\eps\be_1}\de_{\ga_2\be_2}-
\de_{kq_2}\de_{p_2q_1}\de_{\eps\be_2}\de_{\ga_2\be_1})\right.\nn\\&&
-\de_{k'p_2}\de_{\nu\ga_2} (\de_{kq_2}\de_{p_1q_1}\de_{\eps\be_2}\de_{\ga_1\be_1}-
\de_{kq_1}\de_{p_1q_2}\de_{\eps\be_1}\de_{\ga_1\be_2})
+\left.\Theta(\tau)\de_{kk'}\de_{\eps\nu}\de_{s_is_j}\right] 
\eea

where $k,k'=k_1,k_2$.
Using the expressions  of various propagators, it  can  be shown  after some
algebra that
\bea
A^{(2)}_{S_iS_j}&=&-{J^2\over 2 }\int\ d\tau\sum_{a,b}\left\{ tr{\s^a\over 2}{\s^b\over 2}<u_i|S^aS^b|u_j>
\left[ G^2(\tau)+{2\over l} G(\tau)\veps(\tau)\right]\right. \nn\\
&+&{1\over l}G(\tau)~\left( \de_{p_2q_2}\de_{\ga_2\be_2} \left[{\s^a\over 2},{\s^b\over 2}\right]_{\be_1\ga_1}
+\de_{p_1q_1}\de_{\ga_1\be_1} \left[{\s^a\over 2},{\s^b\over 2}\right]_{\be_2\ga_2}\right.
\label{a2}\\&& \left.\left.- \de_{p_2q_1}\de_{\ga_2\be_1} 
\left[{\s^a\over 2},{\s^b\over 2}\right]_{\be_2\ga_1}-
 \de_{p_1q_2}\de_{\ga_1\be_2} \left[{\s^a\over 2},{\s^b\over 2}\right]_{\be_1\ga_2}\right)
{\cal T}\la u_i|S^a(\tau)S^b(0)|u_j\ra 
\right\}\nn
\eea
We are only focusing on terms leading to infrared divergences which will renormalize the Kondo coupling constant.
Notice  that the first line of (\ref{a2}) do not give any logarithm contributions.
The leading contribution at large $l$ can be then computed straightforwardly in all cases.
We can establish
\bea
&&A^{(2)}_{11}=-{J^2\over 2\pi v_F l}\ln lc=-{\pi v_F \lam^2\over 2 l}\ln lc~,\\
&&A^{(2)}_{12}=A^{(2)}_{13}={J^2\over 2\pi v_F l}\ln lc={\pi v_F \lam^2\over 2 l}\ln lc~,\\
&&A^{(2)}_{14}=-A^{(2)}_{15}={J^2\over 2\pi v_F l}\ln lc={\pi v_F \lam^2\over 2 l}\ln lc~,\\
&&A^{(2)}_{24}=A^{(2)}_{34}=-A^{(2)}_{25}=-A^{(2)}_{35}=-{J^2\over 4\pi v_F l}\ln lc=-{\pi v_F \lam^2\over 4 l}\ln lc~.
\eea
All other matrix elements do not give any logarithm divergences.
From these results, it is then easy to compute the matrix elements defined in 
(\ref{degpert}) and to show that the Kondo coupling constant
in the Eq. (\ref{matrix}) gets perfectly renormalized namely $J/l$ has to be replaced by
$\pi v_F(\lam+\lam^2\ln[lc])/l$.


\end{document}